\documentclass[oldversion]{aa}

\usepackage{graphicx}
\usepackage{natbib}

\bibpunct{(}{)}{;}{a}{}{,}
\usepackage{txfonts}
\usepackage{subfigure}

\begin{document}

  \title{Multiwavelength campaign on Mrk 509}

   \subtitle{III. The 600 ks RGS spectrum: unravelling the inner region of an AGN}

   \author{R.G. Detmers\inst{1,2}
          \and
          J.S. Kaastra\inst{1,2}
          \and
          K. Steenbrugge\inst{3,4}
           \and
           J. Ebrero \inst{1}
          \and
          G. A. Kriss\inst{5,6}
          \and
          N. Arav\inst{7}
          \and
          E. Behar\inst{8}
          \and
          E. Costantini\inst{1}
          \and
          G. Branduardi-Raymont\inst{9}
          \and
          M. Mehdipour\inst{9}
          \and
          S. Bianchi\inst{10}
          \and
          M. Cappi\inst{11}
          \and
          P. Petrucci\inst{12}
          \and
          G. Ponti\inst{13}
          \and
          C. Pinto\inst{1}
          \and
          E.M. Ratti\inst{1}
          \and
          T. Holczer\inst{8}
	  }

   \offprints{R.G. Detmers}

   \institute{SRON Netherlands Institute for Space Research, Sorbonnelaan 2, 3584 CA Utrecht, The Netherlands \email{r.g.detmers@sron.nl}
    \and
    Astronomical Institute, University of Utrecht, Postbus 80000, 3508 TA Utrecht, The Netherlands
    \and
    Instituto de Astronom\'ia, Universidad Cat\'olica del Norte, Avenida Angamos 0610, Casilla 1280, Antofagasta, Chile 
    \and
    Department of Physics, University of Oxford, Keble Road, Oxford OX1 3RH, UK
    \and
    Space Telescope Science Institute, 3700 San Martin Drive, Baltimore, MD 21218, USA
    \and
    Department of Physics \& Astronomy, The Johns Hopkins University, Baltimore, MD, 21218, USA
    \and
    Department of Physics, Virginia Tech, Blacksburg, VA 24061, USA
    \and
    Department of Physics, Technion, Haifa 32000, Israel
    \and
    Mullard Space Science Laboratory, University College London, Holmbury St. Mary, Dorking, Surrey, RH5 6NT, UK 
    \and
    Dipartimento di Fisica, Universita degli Studi Roma Tre, via della Vasca Navale 84, I-00146 Roma, Italy 
    \and
    INAF-IASF Bologna, Via Gobetti 101, I-40129 Bologna, Italy 
    \and
    UJF-Grenoble 1 / CNRS-INSU, Institut de PlanŽtologie et d?Astrophysique de Grenoble (IPAG) UMR 5274, Grenoble, F-38041, France.
    \and
    School of Physics and Astronomy, University of Southampton, Highfield, Southampton SO17 1BJ, UK    }     

   \date{}

% \abstract{}{}{}{}{} 
% 5 {} token are mandatory
 
\abstract{We present the results of our 600 ks RGS observation as part of the multiwavelength campaign on Mrk 509. The very high quality of the spectrum allows us to investigate the ionized outflow with an unprecedented accuracy due to the long exposure and the use of the RGS multipointing mode. We detect multiple absorption lines from the interstellar medium and from the ionized absorber in Mrk 509. A number of emission components are also detected, including broad emission lines consistent with an origin in the broad line region, the narrow \ion{O}{vii} forbidden emission line and also (narrow) radiative recombination continua. The ionized absorber consists of two velocity components ($v$ = $-$ 13 $\pm$ 11 km s$^{-1}$ and $v$ = $-$ 319 $\pm$ 14 km s$^{-1}$), which both are consistent with earlier results, including UV data. There is another tentative component outflowing at high velocity, $-$ 770 $\pm$ 109 km $\mathrm{s^{-1}}$, which is only seen in a few highly ionized absorption lines. The outflow shows discrete ionization components, spanning four orders of magnitude in ionization parameter. Due to the excellent statistics of our spectrum, we demonstrate for the first time that the outflow in Mrk 509 in the important range of $\log\xi$ between 1$-$3 cannot be described by a smooth, continuous absorption measure distribution, but instead shows two strong, discrete peaks. At the highest and lowest ionization parameters we cannot differentiate smooth and discrete components.} 

   \maketitle
%
%________________________________________________________________

\section{Introduction}			\label{intro}

One of the main reasons to study active galactic nuclei (AGN) is to learn about feedback from the AGN to the galaxy and its direct environments. 
Feedback is a combination of enrichment (the spreading of elements into the interstellar and inter galactic media (ISM and IGM), momentum feedback (due to winds), and direct kinetic feedback (i.e. energy ejection into the ISM or IGM by jets).
From recent observations on cooling clusters of galaxies \citep[see e.g.][for an overview]{McNamara07}, as well as from recent insights into galaxy and AGN co-evolution \citep{DiMatteo05,Elvis06,Bower09,Fabian10}, it has become clear that feedback from AGN is a crucial ingredient  the evolution of galaxies and clusters of galaxies. This is also seen in the so-called M-$\sigma$ relation, which links the velocity dispersion of stars in the bulge to the mass of the Super-Massive Black Hole (SMBH) \citep{Ferrarese00,Gebhardt00}. While we have a reasonable qualitative understanding of the feedback from relativistic jets \citep[as observed in clusters of galaxies, see e.g.][]{Fabian03}, we still lack a quantitative picture of the feedback of the AGN on the galaxy and on its surroundings. \\
There is a broad ongoing effort to improve this, and recent work on broad absorption line (BAL) quasars shows that the mass outflow rates in these systems are 100s of solar masses per year and the kinetic luminosity involved is a few percent of the total bolometric luminosity \citep{Moe09,Dunn10}. 
There have also been indications that some AGN harbor a highly ionized, massive, ultra-fast outflow, with velocities reaching up to 60\,000 km s$^{-1}$ \citep{Reeves03,Pounds09,Ponti09,Tombesi10a,Tombesi10b}. These outflows are hard to detect, however, and appear to be variable (because they are only present in some observations of a single source). These are extreme cases of outflows that are present in only a fraction of the total number of AGN. Whether feedback from less extreme outflows, such as those that are present in about 50 $\%$ of the local Seyfert 1 galaxies is also important, remains an unsolved question.

If we can establish the impact that the outflow has on the galaxy in these local Seyfert 1 galaxies, this should allow us to extend the feedback estimates that we obtain to higher redshifts to the more powerful AGN, which we are unable to investigate with the current generation of X-ray grating spectrometers. 
However we first need to deal with the two main uncertainties concerning the outflows. These are the geometry of the inner region of an AGN and the location or origin of the outflow \citep[see e.g.][]{Murray97,Krolik01,Gaskell07}. Earlier work has placed the outflow at various distances, and also the estimates for feedback can vary wildly \citep[see e.g.][for some examples]{Behar03,Blustin05,Krongold07,Detmers08}. Therefore answering these two questions is the main goal of the Mrk 509 multiwavelength campaign. 

Multiwavelength campaigns on AGN are crucial for gaining a complete understanding of the inner regions of these sources. Earlier multiwavelength campaigns focused mainly on abundance studies of the outflow \citep[see Mrk 279,][]{Arav07} or on determining the outflow structure and location by combining UV and X-ray data e.g. NGC\,5548 \citep{Steenbrugge05}; NGC\,3783 \citep{Netzer03,Gabel03}.
Our dedicated multiwavelength campaign on Mrk 509 is much more extensive than previous attempts. Our more intensive observations are ideal for locating the outflow, using the variability of the source and response of the ionized gas to determine its location \citep[the use of variability to locate gas has been very successfully used in reverberation mapping of the BLR, see e.g.][for an overview of the method and for the latest results]{Peterson00,Denney10}.  

Apart from the location and kinematics, one of the other important questions regarding the outflow is what the ionization structure is. Earlier studies have reported different results. The outflow in NGC 5548 appears to be a continuous distribution of column density vs. log $\xi$ \citep{Steenbrugge05}. NGC 3783, on the other hand, shows different separate ionization components, all in pressure equilibrium \citep{Krongold03}. In Mrk 279 the situation appears to be more complex, because a nonmonotonous, continuous distribution provides the best description to the data \citep{Costantini07a}. 
Recently, \citet{Holczer07} and \citet{Behar09} have shown that for most local Seyfert 1 galaxies with an outflow, a continuous distribution of column density vs. ionization parameter is the best description of the data. What they also show is that there are distinct $\xi$ values where no is gas present. They interpret these gaps as thermal instabilities that cause the gas to rapidly cool or heat and then shift to other ionization parameters. 
What is clear from these studies is that there is no single model that describes all the observed outflows. High$-$quality, high$-$resolution spectra of the outflows are crucial for investigating the structure, since it can be the case (as in Mrk 279) that some components of the outflow have very low column density, which would otherwise escape detection.      

Mrk 509 is one of the best studied local AGN, and due to its large luminosity ($L(1-1000\,\mathrm{Ryd})$ = 3.2 $\times$ 10$^{38}$ W), it is also considered to be one of the closest QSO/Seyfert 1 hybrids. Earlier work on the outflow in the X-ray regime has revealed that it consists of a wide range of ionization components, but lacks the very high and also very low ionized gas \citep[weak Fe UTA and no \ion{Si}{xiv} Ly$\alpha$,][]{Yaqoob03}. The outflow has been described using three ionization components, each with a different outflow velocity \citep{Smith07}; however, the exact outflow velocities differ between different publications, most likely due to a limited signal$-$to$-$noise ratio. \citet{Detmers10} have analyzed three archival observations of Mrk 509 with \textit{XMM-Newton}. They also find three components for the outflow, although with different velocities than \citet{Smith07}. Including the EPIC-pn data and improving the relative calibration between RGS and EPIC-pn achieved increased sensitivity. With this improvement they were able to detect variability in the highest ionization component, constraining the distance of that component to within 0.5 pc of the central source. Another point of interest is that there have been indications of an ultra-high velocity outflow as seen through the \ion{Fe}{K} line \citep{Cappi09,Ponti09}. This outflow could make a potentially large contribution to feedback, as the velocity is very high, although it appears to be transient \citep{Ponti09}. 
    
This work is the third in a series of papers analyzing the very deep and broad multiwavelength campaign on Mrk 509. The complete campaign details are presented in \citet{Kaastra2011a}, hereafter Paper I. Here we present the main results obtained from the stacked 600 ks \textit{XMM-Newton} RGS spectrum.
With this spectrum, we are able to characterize the properties of the ionized outflow in great detail (velocities, ionization states, column densities, density profile, etc.). Other features detected in the spectrum ( emission lines, Galactic absorption, etc.) will not be discussed in detail here. Because different physics are involved, we will discuss them in future papers in this series.  

This article is organized as follows. Section \ref{model} briefly describes the data reduction for obtaining the stacked spectrum, and we show the spectral models that we use to describe the data. The spectral analysis and the results are presented in Sect. \ref{spectral}. We discuss our results in Sect. \ref{discussion} and present our conclusions in Sect. \ref{conclusions}.  

\section{Data reduction and modeling}			\label{model}

\subsection{Data reduction}

\begin{figure}[htbp]
   \includegraphics[angle= -90,width=9cm]{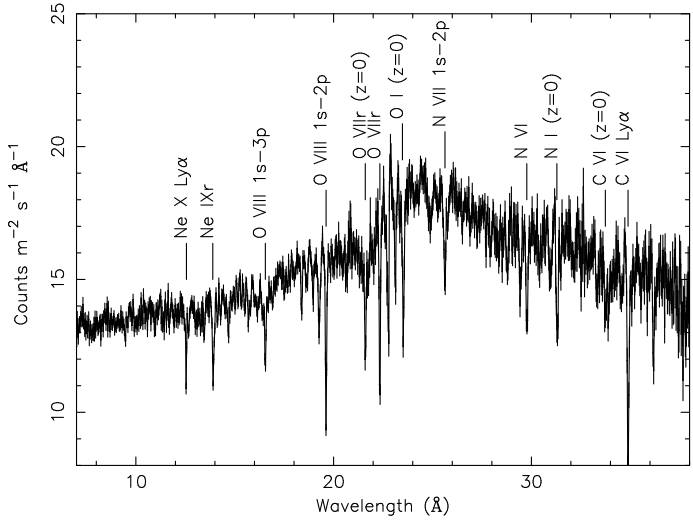}
   \caption{\label{fig:fluxed}
        The fluxed stacked RGS spectrum in the 7 - 38 $\AA$ range. The strongest lines are indicated and the \ion{O}{i} ISM edge can be clearly seen around 23 $\AA$.}
\end{figure}

The RGS data reduction used here is much more complex than the standard pipeline processing using the \textit{XMM-Newton} Science Analysis System (SAS), the main reasons among others being the use of the RGS multi-pointing mode, a variable source and a nonstandard procedure of filtering for bad pixels.
The full details of the data reduction and all the necessary steps are found in \citet{Kaastra2011b}, hereafter paper II. In short, we used the SAS 9.0 software package to reduce all the individual observations. Then we created a fluxed spectrum for each observation and stacked those taking the effects of the multi-pointing mode into account. This way a fluxed RGS spectrum was created by stacking both RGS 1 and 2 and both spectral orders. We used this fluxed spectrum for fitting our data.  
Figure \ref{fig:fluxed} shows the full fluxed spectrum with some of the strongest lines indicated. 

\subsection{Setup}			\label{setup}

We used data between 7 and 38 $\AA$ when fitting the RGS spectrum. As the fluxed spectrum consists of both the RGS 1 and 2 data in both spectral orders, we binned the data between 7 and 38 $\AA$ in 0.01 $\AA$ bins. The average number of counts per 0.01 $\AA$ bin is approximately 900 (i.e. a signal$-$to$-$noise$-$ratio of about 30).  
We therefore used $\chi^{2}$ statistics when fitting the spectrum. All errors are given for $\Delta$$\chi^{2}$ = 1. We used the SPEX 2.03.00\footnote{see http://www.sron.nl/spex} spectral fitting package to fit the spectrum. We updated the wavelengths of some important transitions for our study (see Appendix \ref{sect:app}).

We constructed the spectral energy distribution (SED) for Mrk 509, using the EPIC-pn and OM data to obtain the necessary flux points for the \textit{XMM-Newton} observations and extending it with other data. This SED is an average of the Mrk 509 SED during the time of observations (roughly two months time). The full procedure on how the SED was derived can be found in \citet{Kaastra2011a}. 
The ionization balance calculations needed for our spectral modeling (the \textit{xabs} components, see Sect. \ref{sect:outflow}) were based on this SED and performed using version C08.00 of Cloudy\footnote{http://www.nublado.org/}\citep{Ferland98} with \citet{Lodders09} abundances. 

\subsection{Spectral models}				\label{spectral_models}

The unprecedented quality of the spectrum requires a rather complex spectral model to be described accurately. We describe each component in more detail in separate sections, but we give a short overview of the total model here. 

We model the continuum with a spline (see Fig. \ref{fig:spline_model}). The main reason for doing so is that a spline can accurately describe the (complex) continuum shape without having to make any physical assumptions about the origin of the shape of the continuum (powerlaw, blackbody, Comptonization, or reflection, for example). We use a redshift $z$ = 0.03450, which combines the cosmological redshift \citep{Huchra93} with the orbital velocity of the Earth around the Sun, which is not corrected for in the standard \textit{XMM-Newton} analysis \citep[see][]{Kaastra2011a}. Galactic absorption \citep[N$_{\mathrm{H}}$ = 4.44 $\times$ 10$^{24}$ m$^{-2}$,][]{Murphy96} is also taken into account. We use three distinct phases for the Galactic ISM absorption, a neutral ($kT$ = 0.5 eV) phase, a warm ($kT$ = 4.5 eV) slightly ionized phase, and a hot ($kT$ = 140 eV) highly ionized phase \citep{Pinto10}. Additionally we model the neutral oxygen and iron edges of the ISM by including a dusty component. Details about the Galactic foreground absorption are given by Pinto et al. (in prep).
 
The ionized outflow is modeled with three different models, each with multiple (two or three) velocity components to account for the separate outflow velocities observed. All models take a wide range of ionization states into account. These models are described in more detail in Sect. \ref{sect:outflow}.
We also included eleven broad and narrow emission lines, which are modeled with a Gaussian line profile. Radiative recombination continua (RRCs) are also included using an ad-hoc model that takes the characteristic shape of these features into account .

\section{Spectral analysis}	\label{spectral}

\subsection{Continuum, local absorption, and emission features}		

\begin{figure}[htbp]
   \includegraphics[angle= -90,width=9cm]{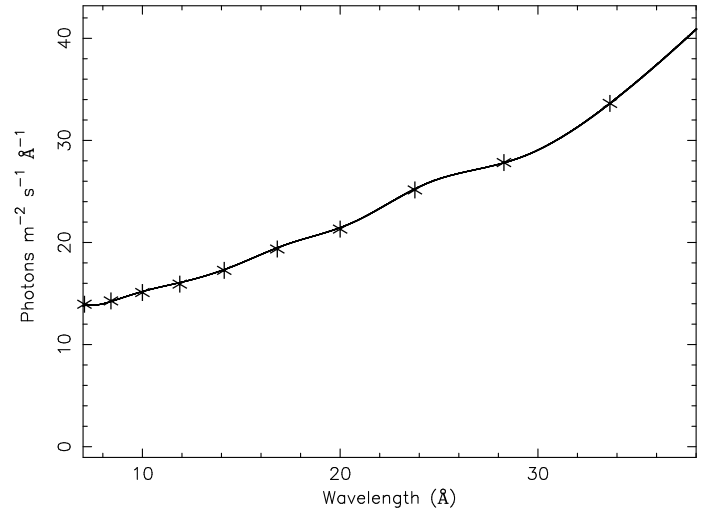}
   \caption{\label{fig:spline_model}
        The unabsorbed spline continuum model used for the Mrk 509 observations. }
\end{figure}

The continuum is modeled with a spline with a logarithmic spacing of 0.075 between 5 and 40 $\AA$. We show the spline model in Fig. \ref{fig:spline_model} and in Table \ref{tab:spline}. The continuum is smooth, so the spline does not mimic any broad line emission features. The softening of the spectrum at longer wavelengths can be seen clearly.   

\begin{table}
\caption{Spline continuum parameters.}             % title of Table
\label{tab:spline}      % is used to refer this table in the text
\centering                          % used for centering table
\begin{tabular}{l@{\,}c}        % centered columns (6 columns)
\hline\hline                 % inserts double horizontal lines
Wavelength & Flux  \\
($\AA$)         & (photons m$^{-2}$ s$^{-1}$ $\AA^{-1}$)\\
\hline                        % inserts single horizontal line
 5.00 & 0   \\
 5.95 & 0.7 $\pm$ 0.6  \\
 7.07 & 13.95 $\pm$ 0.09 \\
 8.41 & 14.28 $\pm$ 0.06 \\
 10.00 & 15.10 $\pm$ 0.04 \\
 11.89 & 15.92 $\pm$ 0.09 \\
 14.14 & 17.27 $\pm$ 0.09 \\
 16.82 & 19.36 $\pm$ 0.16 \\
 20.00 & 21.31 $\pm$ 0.11 \\
 23.78 & 25.15 $\pm$ 0.08\\
 28.28 & 27.83 $\pm$ 0.11\\
 33.64 & 33.62 $\pm$ 0.19 \\
 40.00 & 9.80 $\pm$ 0.18 \\
\hline
\end{tabular}
\end{table}

The neutral Galactic absorption is responsible for the narrow \ion{O}{i} (23.5 $\AA$) and \ion{N}{i} (31.3 $\AA$) absorption lines. To fit the Galactic \ion{O}{ii} absorption line we add a slightly ionized component with a temperature of 4.5 eV and with a column density that is 4\,$\%$ of the cold, neutral gas (Pinto et al., in prep). To properly model the \ion{O}{i} edge, we take the effects of depletion into dust into account. This same depletion is also responsible for the lack of a strong neutral iron absorption blend around 17.5 $\AA$. We use dust that consists of pyroxene and hematite.

The sightline to Mrk 509 passes through a high-velocity cloud located in the Galactic halo \citep[see e.g.][]{Sembach95}. The hot gas as seen in the \ion{C}{vi}, \ion{O}{vii}, \ion{O}{viii}, and \ion{Ne}{ix} absorption lines at $z$ = 0 could either be from ISM absorption in our Milky Way or from this high-velocity cloud (which has an LSR velocity of around $-$ 250 km s$^{-1}$). 
 A more detailed multiwavelength analysis and comparison of all these local components will be presented in a follow-up paper (Pinto et al. ,in prep).

\begin{figure*}[tbp]
\centering
\hbox{
   \includegraphics[angle= -90,width=9cm]{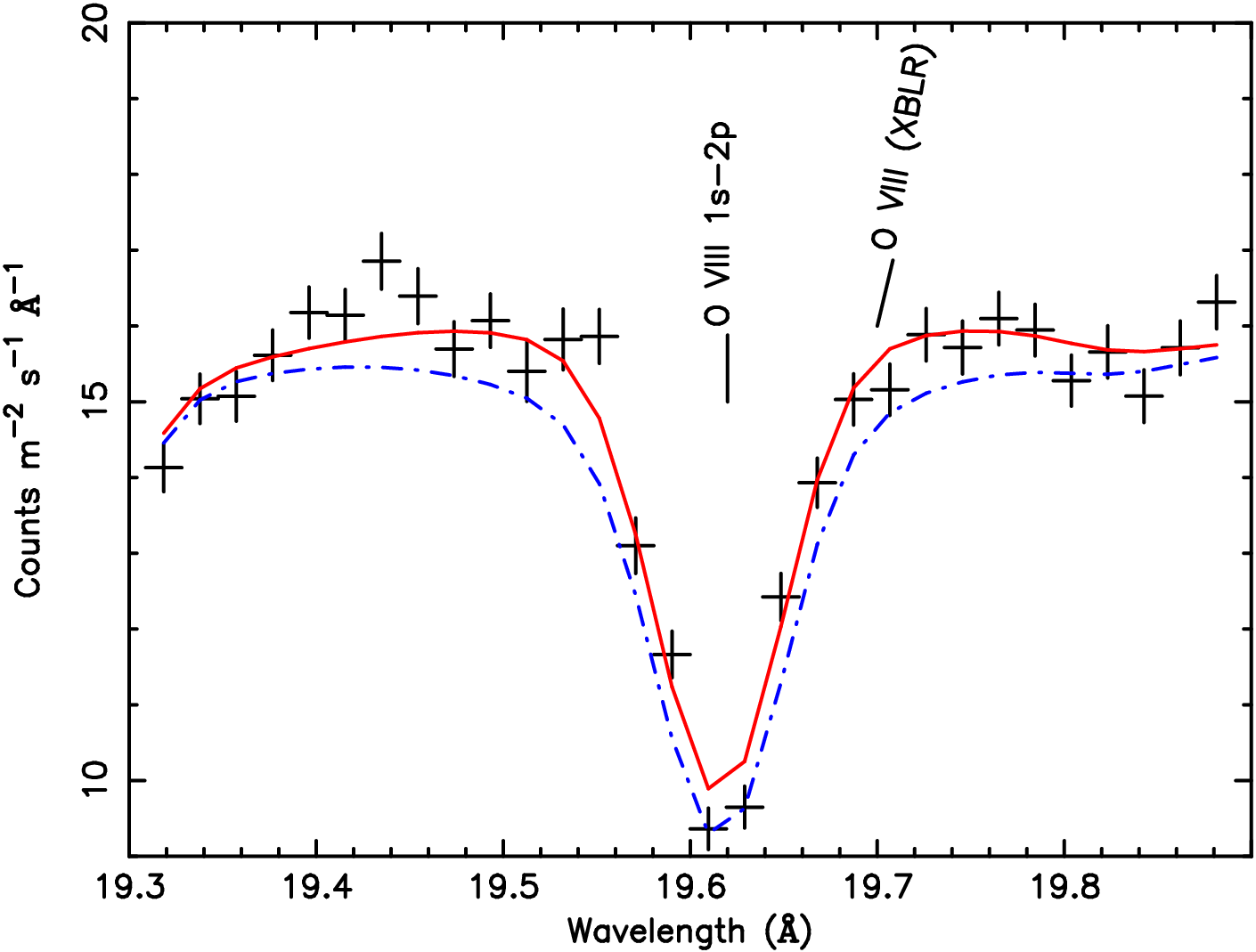}
      \includegraphics[angle= -90,width=9cm]{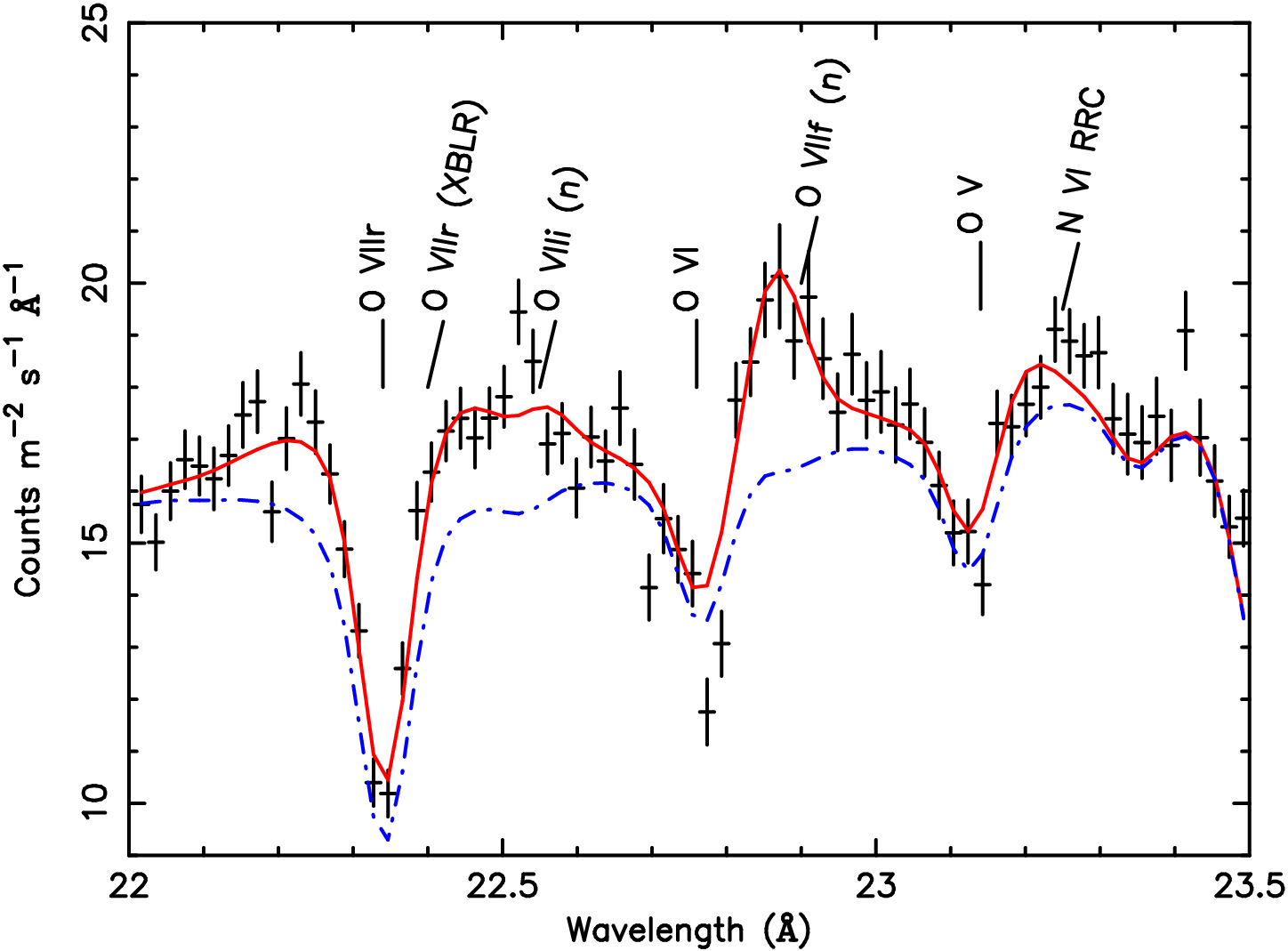}
}
   \caption{\label{fig:broad_lines}
   Broad and narrow X-ray emission lines of \ion{O}{viii} Ly\,$\alpha$ (left) and \ion{O}{vii} (right).
   The model without any broad lines is shown as the dashed-dotted line, while the model (model 2) with the lines and RRC included (marked as XBLR and n) is shown as a solid line. }
\end{figure*}

The spectrum shows some emission lines, most of which are broadened (Table \ref{tab:broad_lines}). In this table we also show the change in $\chi^{2}$ when the line is omitted from the model. 
The broad emission lines are visible as excesses on both sides of the corresponding absorption lines (see Fig. \ref{fig:broad_lines}). 
In our modeling we fix the width of the lines to an FWHM of  4200 $\pm$ 200 km s$^{-1}$ as measured for the Balmer lines simultaneously by the OM optical grism \citep{Mehdipour11}. We assume that these lines originate in the BLR \citep{Costantini07a}.

\begin{table}
\caption{Broad emission line parameters for the combined spectrum, where fluxes are corrected for Galactic and intrinsic absorption.}             % title of Table
\label{tab:broad_lines}      % is used to refer this table in the text
\centering                          % used for centering table
\begin{tabular}{l@{\,}c@{\,}c@{\,}c}        % centered columns (4 columns)
\hline\hline                 % inserts double horizontal lines
Line & Wavelength  & Flux &  $\Delta$$\chi^{2}$\\ % table heading 
         & ($\AA$)          & (ph m$^{-2}$ s$^{-1}$)   & \\
\hline                        % inserts single horizontal line
   \ion{O}{vii} r & 21.602  &  1.00$\pm$0.09  & 82\\      % inserting body of the table
   \ion{O}{vii} i & 21.802  &  $<$ 0.56 & 0\\      % inserting body of the table
   \ion{O}{vii} f & 22.101  &  $<$ 0.7 & 0\\      % inserting body of the table
   \ion{O}{viii} Ly$\alpha$ & 18.967 &  0.42$\pm$0.04  & 46\\
   \ion{N}{vii} Ly$\alpha$  & 24.780 &  0.09$\pm$0.06  & 1\\
   \ion{C}{vi} Ly$\alpha$  & 33.736 &  0.25$\pm$0.10  & 3\\
   \ion{Ne}{ix} r  & 13.447 &  0.26$\pm$0.03 & 40\\
\hline
\multicolumn{4}{l}{$^1$ All the wavelengths were fixed to the laboratory wavelength.}\\
\end{tabular}
\end{table}

We also detect the narrow \ion{O}{vii} f emission line with an absorption$-$corrected flux of 0.46 $\pm$ 0.06 ph m$^{-2}$ s$^{-1}$, as well as a weaker intercombination line with a flux of 0.13 $\pm$ 0.04 ph m$^{-2}$ s$^{-1}$ and an \ion{Ne}{ix} f emission line with a flux of 0.09 $\pm$ 0.02 ph m$^{-2}$ s$^{-1}$. The narrow resonance line cannot be separated from the absorption line, so we have assumed a ratio of 3:1 for photoionized gas for the forbidden$-$to$-$recombination line ratio \citep{Porquet00}.
We do not detect any other narrow emission lines. In our spectral modeling we adopt an FHWM of 610 km s$^{-1}$ for these narrow emission lines, based on the width of the optical [\ion{O}{iii}] emission line \citep{Phillips83}. We assume here that these lines are produced in the NLR \citep{Guainazzi07}. 
Radiative recombination continua (RRC) have been detected in other Seyfert 1 spectra \citep[see e.g. NGC 3783 or Mrk 279,][]{Kaspi02,Costantini07a}, so we checked whether they are present in Mrk 509.  The RRCs are weak and hard to detect (Table \ref{tab:rrc}). We modeled them using the \textit{rrc} model of SPEX and obtain a temperature of 5.4 $\pm$ 2.5 eV for the RRC. This low temperature is a direct indication of photoionized gas. 

\begin{table}
\caption{RRC parameters.}             % title of Table
\label{tab:rrc}      % is used to refer this table in the text
\centering                          % used for centering table
\begin{tabular}{l@{\,}c@{\,}c}        % centered columns (4 columns)
\hline\hline                 % inserts double horizontal lines
Ion & Wavelength & Flux    \\ % table heading 
         & ($\AA$)        & ph m$^{-2}$ s$^{-1}$ \\
\hline                        % inserts single horizontal line
   \ion{O}{vii}   & 16.78 & 0.08 $\pm$ 0.04\\      % inserting body of the table
   \ion{O}{viii}  & 14.23 & $<$ 0.01\\      % inserting body of the table
   \ion{C}{v}     & 31.63 & 0.18 $\pm$ 0.14\\      % inserting body of the table
   \ion{C}{vi}    & 25.30 & 0.16 $\pm$ 0.08\\      % inserting body of the table
   \ion{N}{vi}    & 22.46 & 0.41 $\pm$ 0.18\\      % inserting body of the table
   \ion{N}{vii}   & 18.59 & $<$ 0.08 \\      % inserting body of the table
   \ion{Ne}{ix}  & 10.37  & 0.06 $\pm$ 0.03\\      % inserting body of the table
   \ion{Ne}{x}   & 9.10 & $<$ 0.01\\      % inserting body of the table
\hline
\end{tabular}
\end{table}

\subsection{Ionized outflow}	\label{sect:outflow}

We use three different models for characterizing the ionized outflow. We describe each model separately and then compare the results obtained by all three models. By comparing the results for the different models for the ionized outflow, we can investigate its ionization structure and density profile.  

The first model (Model 1 from here on, Table \ref{tab:slab}) contains two \textit{slab} components. The \textit{slab} model of SPEX calculates the transmission of a slab of material with arbitrary ionic column densities, outflow velocity $v$, and r.m.s. velocity broadening $\sigma$ as free parameters.  We assume a covering factor of unity for both components. The \textit{slab} components have a different outflow velocity and velocity broadening. However, we assign each ion to one of the components based on the ionization parameter $\xi$, which is defined in the following way:
\begin{equation}		\label{eq:xi}
\centering
      \xi = \frac{L}{n r^{2}}, 
\end{equation}
where $L$ is the 1 $-$ 1000 Rydberg luminosity, $n$ the hydrogen number density of the gas and, $r$ the distance from the ionizing source. For Mrk 509 we obtain an ionizing luminosity of $L$ = 3.2 $\times$ 10$^{38}$ W from the SED.
All ions with log $\xi$ $<$ 2.1 are assigned to the first component, while the others are in the second component. The division is based on a preliminary fit where the outflow velocity of individual ions was a free parameter. The first component has an outflow velocity $v$ of $-$ 57 $\pm$ 8 km s$^{-1}$ and a velocity broadening $\sigma$ = 158 $\pm$ 5 km s$^{-1}$, while the second component has an outflow velocity $v$ = $-$ 254 $\pm$ 40 km s$^{-1}$ and $\sigma$ = 133 $\pm$ 30 km s$^{-1}$. The ionization parameters are given in Table \ref{tab:slab} and are those for which the ion fraction peaks for that ion. The fit gives $\chi^{2}$ = 3643 for 3109 degrees of freedom (\textit{dof}). We also show in Table \ref{tab:slab} the best-fit velocities for individual ions when we leave the velocity free compared to the velocities of all the other ions. 

The second model (Model 2 from here on, Table \ref{tab:slab}) is an extension of Model 1. Instead of separating the ions according to their ionization parameter, we include all ions for both velocity components. Additionally we add a third velocity component to account for a high$-$velocity ($v$ = $-$ 770 km $\mathrm{s^{-1}}$) component \citep[tentatively detected in the Chandra HETGS spectrum, see][]{Yaqoob03} mainly to get the appropriate line centroid for the \ion{Fe}{xxi} and \ion{Mg}{xi} absorption lines. This model is more realistic than Model 1 because it assumes a multivelocity structure for every absorption line, which is consistent with what has been observed in earlier UV observations of Mrk 509 \citep{Kriss00,Kraemer03}. The fit gives $\chi^{2}$ = 3589 for 3070 \textit{dof}.

The third model (Model 3 from here on, Table \ref{tab:xabs}) is based on multiple photoionization components (\textit{xabs}). In each \textit{xabs} component the ionic column densities are related through the ionization parameter $\xi$. 
Free parameters are the hydrogen column density $N_{\mathrm{H}}$, ionization parameter $\xi$, r.m.s. velocity width $\sigma$, and outflow velocity $v$. Based on the results of Model 2, we start with one \textit{xabs} component for each of the two main velocity components detected. We add extra \textit{xabs} components until the fit no longer improves.   
A fit with only one \textit{xabs} component for each velocity results in a fit of $\chi^{2}$ = 4261 for 3145 \textit{dof}.
Adding an additional \textit{xabs} component for each outflow velocity improves the fit by $\Delta$\,$\chi^{2}$ = 264/4 \textit{dof}. We use the same outflow velocity and r.m.s. velocity for the \textit{xabs} components of each velocity component.
If we add a third pair of  \textit{xabs} components we again improve the fit significantly resulting in a further improvement of $\Delta$\,$\chi^{2}$ = 138/4 \textit{dof}.
As a last step, we leave the outflow velocity and the r.m.s. velocity width for each component free. This then improves the fit by $\Delta$\,$\chi^{2}$ = 32/4 \textit{dof}. The best fit we obtain this way has $\chi^{2}$  = 3827 for  3157 \textit{dof}. The results for the final fit are shown in Table \ref{tab:xabs}, however the component with the highest ionization parameter (component E2 in Table \ref{tab:xabs}) shifts to a much higher outflow velocity, namely 492 km s$^{-1}$. We therefore add it to the fast velocity group, so that the slow outflow can now be described properly with two \textit{xabs} components, while the fast outflow is described by four components. We label each component according to its ionization parameter (A to E for increasing $\xi$) with index 1 or 2 for low and high outflow velocity, respectively. 

As a test we also fit a fourth model, the so-called \textit{warm} model, essentially a power-law distribution of \textit{xabs} components. This is similar to the model used by \citet{Steenbrugge05} in NGC 5548 (model D in that paper) and akin to the absorption measure distribution (AMD) method used by \citet{Holczer07}. 
We first defined a range of ionization parameters, between which we fit our model. In our case we fit the model between log $\xi$ = $-$2 and 4. We used a grid of 19 points in order to accurately describe the total AMD. At every grid point a value $f_{i}$ was determined, which is defined as $f_{i}$ = $d\,N_{\mathrm{H}}\,/d\mathrm{log}\xi$. This way we obtained the distribution of $N_{\mathrm{H}}$ versus log $\xi$.  We did this for both velocity components (ignoring the very fast $-$770 km s$^{-1}$ outflow at the moment). The best fit we obtain has $\chi^{2}$ = 3822 for 3120 \textit{dof}.
However due to the correlation between the two warm component parameters (especially the factors $f_{i}$), calculating the exact error on every parameter is very difficult. 

We therefore use the \textit{warm} model only to check whether we have missed any ions in our \textit{slab} fit. With a continuous model like \textit{warm}, weaker lines that otherwise might be hard to detect are taken into account. 
In this way we have identified several ions, which are predicted to be present based on the \textit{warm} model, but were not included in the \textit{slab} fit since they produce only weak lines. These are \ion{Ne}{vii} and \ion{Si}{x} - \ion{Si}{xii}. All these ions, however, have very weak lines in the Mrk 509 spectrum, and the fitted ionic column densities (using the \textit{slab} model) only yield upper limits.
We therefore are confident that Models 2 and 3 are accurate representations of the data.
 
\begin{table*}[!htbp]
\caption{Predicted and measured Ionic column densities for the most important ions of the outflow for model 1 and model 2.}            
\label{tab:slab}      % is used to refer this table in the text
\centering                          % used for centering table
\begin{tabular}{l c c c c c c c c c c c c c}        % centered columns (4 columns)
\hline\hline                 % inserts double horizontal lines
log $\xi^{a}$ & ion$^{b}$ &N$_{\mathrm{ion,obs}}^{c}$ & N$_{\mathrm{ion,pred}}^{d}$ & $\Delta\chi^{e}$ & A$^{f}$ & B$^{g}$ & C$^{h}$ & D$^{i}$ & E$^{j}$ & N$_{\mathrm{ion,obs,v1}}^{k}$ & N$_{\mathrm{ion,obs,v2}}^{l}$ & N$_{\mathrm{ion,obs,v3}}^{m}$ & v$_{ion}^{n}$\\
\hline  
$-$8.50 & \ion{H}{i} & & 1.20 & & 79 & 20 & 1 &	0 & 0 & & &\\
$-$8.50 & \ion{Fe}{i} & $<$ 1.00 & 0.00 & & 0 & 0 & 0 & 0 & 0 & $<$ 0.08 & $<$ 0.08 &\\
$-$8.50 & \ion{Fe}{ii} & $<$ 0.63 & 0.00 & & 0 & 0 & 0 & 0 & 0 & $<$ 0.50 & $<$ 0.32 &\\
$-$2.10 & \ion{Fe}{iii} & $<$ 0.03 & 0.00 & & 100 & 0 & 0 & 0 & 0 & $<$ 0.32 & $<$ 0.13 &\\
$-$1.65 & \ion{H}{ii}   & & 604980 & & 0 & 1 & 8 & 9 & 81 &  & &\\
$-$1.48 & \ion{O}{iii}   & $<$ 0.25 & 0.09 & & 99 & 1 & 0 & 0 & 0 & $<$ 0.16 & $<$ 0.25 &\\
$-$1.45 & \ion{C}{iii}   & & 0.10 & & 97 & 3 & 0 & 0 & 0 &  & &\\
$-$1.42 & \ion{Fe}{iv} & $<$ 0.03 & 0.00 & & 100 & 0 & 0 & 0 & 0 & $<$ 0.05 & $<$ 0.16 &\\
$-$0.95 & \ion{Fe}{v} & $<$ 0.25 & 0.01 & & 99 & 1 & 0 & 0 & 0 & $<$ 0.08 & $<$ 0.10 &\\
$-$0.65 & \ion{O}{iv}   & 0.87 $\pm$ 0.23 & 0.86 & 0.1 & 90 & 10 & 0 & 0 & 0 & 0.72 $\pm$ 0.19 & $<$ 0.03 & &270 $\pm$ 170\\
$-$0.60 & \ion{C}{iv}   & & 0.25 & & 80 & 20 & 0 & 0 & 0 & & &\\
$-$0.55 & \ion{N}{iv}   & 0.17 $\pm$ 0.20 & 0.09 & 0.4 & 85 & 15 & 0 & 0 & 0 & 0.1 $\pm$ 0.3 & $<$ 0.13 &\\
$-$0.55 & \ion{Fe}{vi} & $<$ 0.20 & 0.03 & & 86 & 14 & 0 & 0 & 0 & $<$ 0.16 & $<$ 0.10 &\\
$-$0.05 & \ion{O}{v}     & 1.26 $\pm$ 0.15 & 1.24 & 0.1 & 41 & 59 & 0 & 0 & 0 & 0.47 $\pm$ 0.24 & 1.0 $\pm$ 0.5 & &$-$ 160 $\pm$ 60\\
  0.00 & \ion{N}{v}   & $<$ 0.40 & 0.11 & & 39 & 61 & 0 &	0 & 0 & $<$ 0.01 & $<$ 0.50 &\\
  0.05 & \ion{Fe}{vii}   & $<$ 0.06 & 0.09 & & 28 & 72 & 0 & 0 & 0 & $<$ 0.03 & 0.19 $\pm$ 0.04 &\\
  0.15 & \ion{C}{v}     & 1.8 $\pm$ 0.3 & 2.01 & $-$0.7 & 19 & 78 & 3 & 0 & 0 & 2.04 $\pm$ 0.25 & $<$ 0.02 & &$-$ 10 $\pm$ 50\\
  0.42 & \ion{O}{vi}     & 1.78 $\pm$ 0.22 &	 1.76 & 0.1 & 7 & 91 & 2 & 0 & 0 & 1.00 $\pm$ 0.23 & 0.7 $\pm$ 1.8 & &$-$ 120 $\pm$ 60\\
  0.75 & \ion{N}{vi}     & 0.62 $\pm$ 0.06 &	 0.62 & 0.0 & 3 & 76 & 21 & 0 & 0 & 0.59 $\pm$ 0.10 & 0.06 $\pm$ 0.10 & &$-$ 10 $\pm$ 50\\
  0.85 & \ion{Fe}{viii}   & $<$ 0.25 & 0.17 & & 3 & 94 & 3 & 0 & 0 & $<$ 0.20 & $<$ 0.04 &\\
  0.92 & \ion{Ar}{ix}   & 0.07 $\pm$ 0.08 & 0.01 & 0.8 & 0 & 97 & 3 & 0 & 0 & $<$ 0.03 & $<$ 0.10 &\\
  1.15 & \ion{O}{vii}     & 8.5 $\pm$ 0.6 & 8.55 &$-$0.1 & 0 & 38 & 60 & 1 & 0 & 8.1 $\pm$ 1.0 & 1.2 $\pm$ 0.3 & &$-$ 80 $\pm$ 20\\
  1.20 & \ion{C}{vi}      & 3.6 $\pm$ 0.3 & 3.53 & 0.4 & 0 & 28 & 63 & 6 & 3 & 3.1 $\pm$ 0.4 & 0.62 $\pm$ 0.11 & &$-$ 50 $\pm$ 20\\
  1.22 & \ion{Ne}{viii}   & 0.3 $\pm$ 0.3 & 0.38 &$-$0.4 & 0 & 34& 66 & 0 & 0 & 0.22 $\pm$ 0.12 & $<$ 0.03 &\\
  1.35 & \ion{Ar}{x} & 0.35 $\pm$ 0.18 & 0.01 & 1.9 & 0 & 34 & 66 & 0 & 0 & $<$ 0.50 & $<$ 0.05 &\\
  1.37 & \ion{Fe}{ix}    & $<$ 1.26 & 0.09 & & 0 & 32& 68 & 0 & 0 & $<$ 0.04 & $<$ 0.03 &\\
  1.60 & \ion{N}{vii}      & 1.51 $\pm$ 0.11 & 1.51 & 0.0 & 0 & 5 & 80 & 10 & 5 & 0.93 $\pm$ 0.11 & 0.55 $\pm$ 0.10 & &$-$ 120 $\pm$ 40\\
  1.65 & \ion{Fe}{x} & 0.22 $\pm$ 0.04 & 0.20 & 0.5 & 0 & 2 & 98 & 0 & 0 & $<$ 0.03 & 0.26 $\pm$ 0.04 & &$-$ 350 $\pm$ 220\\
  1.71 & \ion{Ar}{xi} & $<$ 0.01 & 0.04 & & 0 & 1& 99 & 0 & 0 & $<$ 0.02 & $<$ 0.01 &\\
  1.75 & \ion{Ne}{ix}  & 5.0 $\pm$ 0.6 & 5.00 & 0.0 & 0 & 1 & 92 & 7 & 0 & 2.2 $\pm$ 0.7 & 2.0 $\pm$ 0.7 & 0.32 $\pm$ 0.25 & $-$ 160 $\pm$ 40\\
  1.85 & \ion{Fe}{xi} & 0.13 $\pm$ 0.10& 0.29 & $-$1.6 & 0 & 0 & 100 & 0 & 0 & 0.05 $\pm$ 0.04 & $<$ 0.10 &\\
  1.88 & \ion{O}{viii}   & 21.9 $\pm$ 2.1 & 21.97 & 0.0 & 0 & 1 & 75 & 16 & 8 & 13.5 $\pm$ 2.4 & 5.1 $\pm$1.5 & &$-$ 100 $\pm$ 20\\
  1.97 & \ion{Fe}{xii} & 0.10 $\pm$ 0.07 & 0.26 & $-$2.4 & 0 & 0 & 100 & 0 & 0 & $<$ 0.03 & 0.23 $\pm$ 0.05 &\\
  2.00 & \ion{Ca}{xiii} & $<$ 0.05 & 0.18 & & 0 & 0 & 96 & 4 & 0 & $<$ 0.05 & $<$ 0.02 &\\
  2.04 & \ion{Fe}{xiii} & 0.26 $\pm$ 0.05 & 0.19 & 1.7 & 0 & 0 & 100 & 0 & 0 & 0.28 $\pm$ 0.16 & $<$ 0.06 & &$-$ 20 $\pm$ 190\\
  2.05 & \ion{Ar}{xii} & $<$ 0.00	& 0.06 & & 0 & 0 & 98 & 2 & 0 & $<$ 0.03 & $<$ 0.01 &\\
  2.10 & \ion{Fe}{xiv} & 0.30 $\pm$ 0.05 & 0.11 & 3.7 & 0 & 0 & 100 & 0 & 0 & $<$ 0.06 & $<$ 0.05 & &$-$ 370 $\pm$ 400\\
  2.15 & \ion{Fe}{xv} & $<$ 0.05	& 0.05 & & 0 & 0 & 99 & 1 & 0 & $<$	0.50 & 0.2 $\pm$ 0.4 &\\
  2.20 & \ion{S}{xii} & 0.13 $\pm$ 0.07 & 0.20 & $-$1.1 & 0 & 0 & 95 & 5 & 0 & 0.17 $\pm$ 0.07 & $<$ 0.04 & &10 $\pm$ 110\\
  2.21 & \ion{Fe}{xvi} & 0.17 $\pm$ 0.17 & 0.04 & 0.8 & 0 & 0 & 95 & 5 & 0 & 0.10 $\pm$ 0.04 & $<$ 0.03 & &$-$ 60 $\pm$ 230\\
  2.25 & \ion{Mg}{xi} & 1.7 $\pm$ 0.5 & 1.74 & 0.0 & 0 & 0 & 69 & 29 & 2 & $<$ 0.63 & $<$ 12.6 & 1.78 $\pm$ 1.08 & $-$ 640 $\pm$ 210\\
  2.30 & \ion{Fe}{xvii} & 0.20 $\pm$ 0.04 & 0.27 & $-$1.5 & 0 & 0 & 59 & 41 & 0 & $<$ 0.01 & 0.19 $\pm$ 0.04 & &$-$ 400 $\pm$ 130\\
  2.35 & \ion{Ca}{xiv} & 0.13 $\pm$ 0.03 & 0.13 & 0.0 & 0 & 0 & 71 & 29 & 0 & $<$ 0.04 & 0.13 $\pm$ 0.03 & &$-$ 400 $\pm$ 150\\
  2.42 & \ion{S}{xiii} & 0.24 $\pm$ 0.07 & 0.09 & 2.2 & 0 & 0 & 61 & 39 & 0 & 0.09 $\pm$ 0.14 & 0.16 $\pm$ 0.09 & &$-$ 230 $\pm$ 180\\
  2.42 & \ion{Ne}{x} & 7.9 $\pm$ 2.3 & 7.71 & 0.1 & 0 & 0 & 34 & 38 & 28 & $<$ 0.63 & 9.5 $\pm$ 1.4 & $<$ 0.32 & $-$ 270 $\pm$ 50\\
  2.51 & \ion{Fe}{xviii} & 0.60 $\pm$ 0.07 & 0.45 & 2.0 & 0 & 0 & 11 & 89 & 0 & $<$ 0.08 & 0.43 $\pm$ 0.07 & &$-$ 360 $\pm$ 110\\
  2.60 & \ion{S}{xiv} & 0.12 $\pm$ 0.12 & 0.10 & 0.2 & 0 & 0 & 10 & 86 & 3 & $<$ 0.13 & 0.10 $\pm$ 0.17 &\\
  2.77 & \ion{Fe}{xix} & 0.62 $\pm$ 0.06 & 0.66 & $-$0.7 & 0 & 0 & 1 & 98 & 1 & $<$ 0.50 & 0.71 $\pm$ 0.05 & &$-$ 210 $\pm$ 100\\
  3.01 & \ion{Fe}{xx} & 0.48 $\pm$ 0.08 & 0.41 & 0.8 & 0 & 0 & 0 & 90 & 10 & 0.28 $\pm$ 0.12 & 0.4 $\pm$ 0.3 & &$-$ 680 $\pm$ 300\\
  3.20 & \ion{Fe}{xxi} & 0.28 $\pm$ 0.10 & 0.30 & $-$0.3 & 0 & 0 & 0 & 35 & 65 & $<$ 0.06 & $<$ 0.06 & 0.39 $\pm$ 0.13 & $-$ 800 $\pm$ 220\\
  3.31 & \ion{Fe}{xxii} & $<$ 0.32 & 0.60 & & 0 & 0 & 0 & 3 & 97 & 0.2 $\pm$ 0.3 & $<$ 0.16	 &\\
  3.41 & \ion{Fe}{xxiii} & $<$ 0.40 & 1.39 & & 0 & 0 & 0 & 0 & 100 & $<$ 0.40 & $<$ 0.40 &\\
  3.52 & \ion{Fe}{xxiv} & $<$ 2.51 & 2.83 & & 0 & 0 & 0 & 0 & 100 & $<$ 0.32 & $<$ 0.79 &\\
\hline                                   %inserts single line
\multicolumn{14}{l}{$^a$ Ionization parameter where the ion has its peak concentration in 10$^{-9}$ W m.} \\
\multicolumn{14}{l}{$^b$ Element and ionization degree.} \\
\multicolumn{14}{l}{$^c$ Observed column density in 10$^{20}$ m$^{-2}$ for model 1.} \\
\multicolumn{14}{l}{$^d$ Predicted column density in 10$^{20}$ m$^{-2}$ for model 1.} \\
\multicolumn{14}{l}{$^e$ Difference in $\Delta\chi$ between predicted and observed column density. } \\
\multicolumn{14}{l}{$^f$ Percentage of ionic column density produced by component A (see Table \ref{tab:robtab}). } \\
\multicolumn{14}{l}{$^g$ Percentage of ionic column density produced by component B (see Table \ref{tab:robtab}). } \\
\multicolumn{14}{l}{$^h$ Percentage of ionic column density produced by component C (see Table \ref{tab:robtab}). } \\
\multicolumn{14}{l}{$^i$ Percentage of ionic column density produced by component D (see Table \ref{tab:robtab}). } \\
\multicolumn{14}{l}{$^j$ Percentage of ionic column density produced by component E (see Table \ref{tab:robtab}). } \\
\multicolumn{14}{l}{$^k$ Observed column density in 10$^{20}$ m$^{-2}$ for model 2, velocity component 1 ($v$ = $-$ 13 $\pm$ 11 km s$^{-1}$). The velocity broadening $\sigma$ = 125 $\pm$ 8 km s$^{-1}$.} \\
\multicolumn{14}{l}{$^l$ Observed column density in 10$^{20}$ m$^{-2}$ for model 2, velocity component 2 ($v$ = $-$ 319 $\pm$ 14 km s$^{-1}$). The velocity broadening $\sigma$ = 107 $\pm$ 9 km s$^{-1}$. } \\
\multicolumn{14}{l}{$^m$ Observed column density in 10$^{20}$ m$^{-2}$ for model 2, velocity component 3 ($v$ = $-$ 770 $\pm$ 109  km s$^{-1}$). The velocity broadening $\sigma$ = 160 $\pm$ 120 km s$^{-1}$.} \\
\multicolumn{14}{l}{$^n$ Outflow velocity for the individual ion in km s$^{-1}$. Only ions with a solid measure of the column density are included (i.e. no upper limits).} \\
\end{tabular}
\end{table*} 
 
\begin{table} [htbp]
\caption{Parameters for model 3.}             % title of Table
\label{tab:xabs}      % is used to refer this table in the text
\centering                          % used for centering table
\begin{tabular}{l@{\,}c c c c@{\,}c}        % centered columns (5 columns)
\hline\hline                 % inserts double horizontal lines
Comp & log $\xi^{a}$ & N$_{H}^{b}$ & $\sigma^{c}$ & $v^{d}$ & log U$^{e}$ \\    % table heading 
\hline                        % inserts single horizontal line
B1 & 0.81 $\pm$ 0.07 & 0.8 $\pm$ 0.1 & 124 $\pm$ 20 & 25 $\pm$ 30 & $-$0.73 \\
C1 & 2.03 $\pm$ 0.02 & 2.6 $\pm$ 0.2 & 193 $\pm$ 14 & $-$ 43 $\pm$ 20 & 0.49  \\
\hline
A2 & $-$0.14 $\pm$ 0.13 & 0.4 $\pm$ 0.1 & 79 $\pm$ 26 & $-$ 180 $\pm$ 41 & $-$1.68 \\
C2 & 2.20 $\pm$ 0.02 & 4.4 $\pm$ 0.5 & 29 $\pm$ 6 & $-$ 267 $\pm$ 31 & 0.66 \\
D2 & 2.62 $\pm$ 0.08 & 1.8 $\pm$ 0.5 & 34 $\pm$ 19 & $-$ 254 $\pm$ 35 & 1.08 \\
E2 & 3.26 $\pm$ 0.06 & 6.3 $\pm$ 1.2 & 37 $\pm$ 19 & $-$ 492 $\pm$ 45 & 1.72 \\
\hline
\multicolumn{6}{l}{$^a$ Ionization parameter in 10$^{-9}$ W m.} \\
\multicolumn{6}{l}{$^b$ Column density in units of 10$^{24}$ m$^{-2}$.} \\
\multicolumn{6}{l}{$^c$ r.m.s. velocity broadening in km s$^{-1}$} \\
\multicolumn{6}{l}{$^d$ Outflow velocity in km s$^{-1}$, a negative velocity corresponds to a blueshift.} \\
\multicolumn{6}{l}{$^e$ ionization parameter (as used in UV spectroscopy). } \\
\end{tabular}
\end{table}  
 
\subsection{Spectral fit} 		\label{spectral_fit}

\begin{figure*}[htbp]
\begin{minipage}[b]{1.0\linewidth}
\centering
\includegraphics[width=12cm, angle = -90]{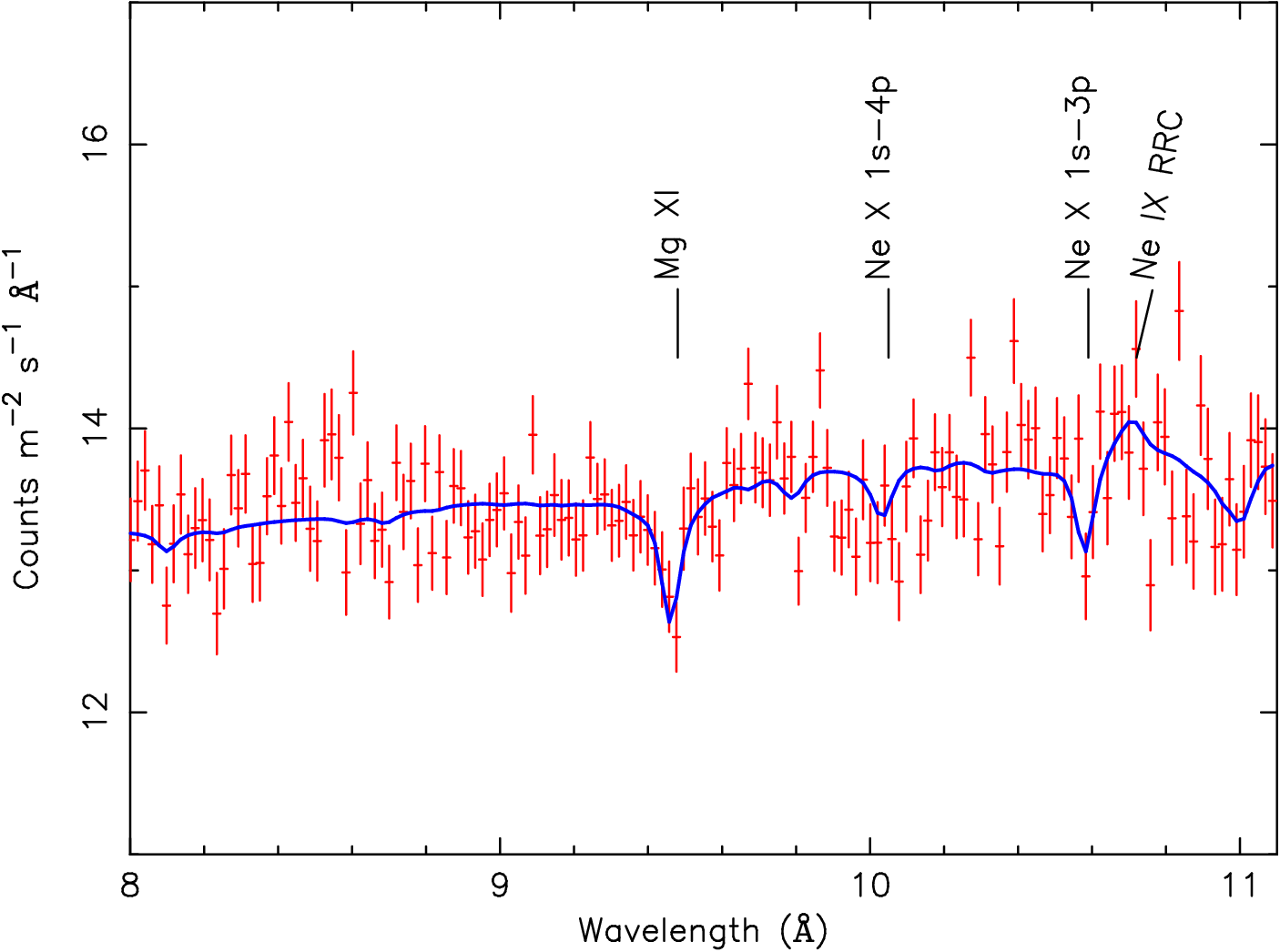}
\end{minipage}
\hspace{2cm}
\begin{minipage}[b]{1.0\linewidth}
\centering
\includegraphics[width=12cm, angle = -90]{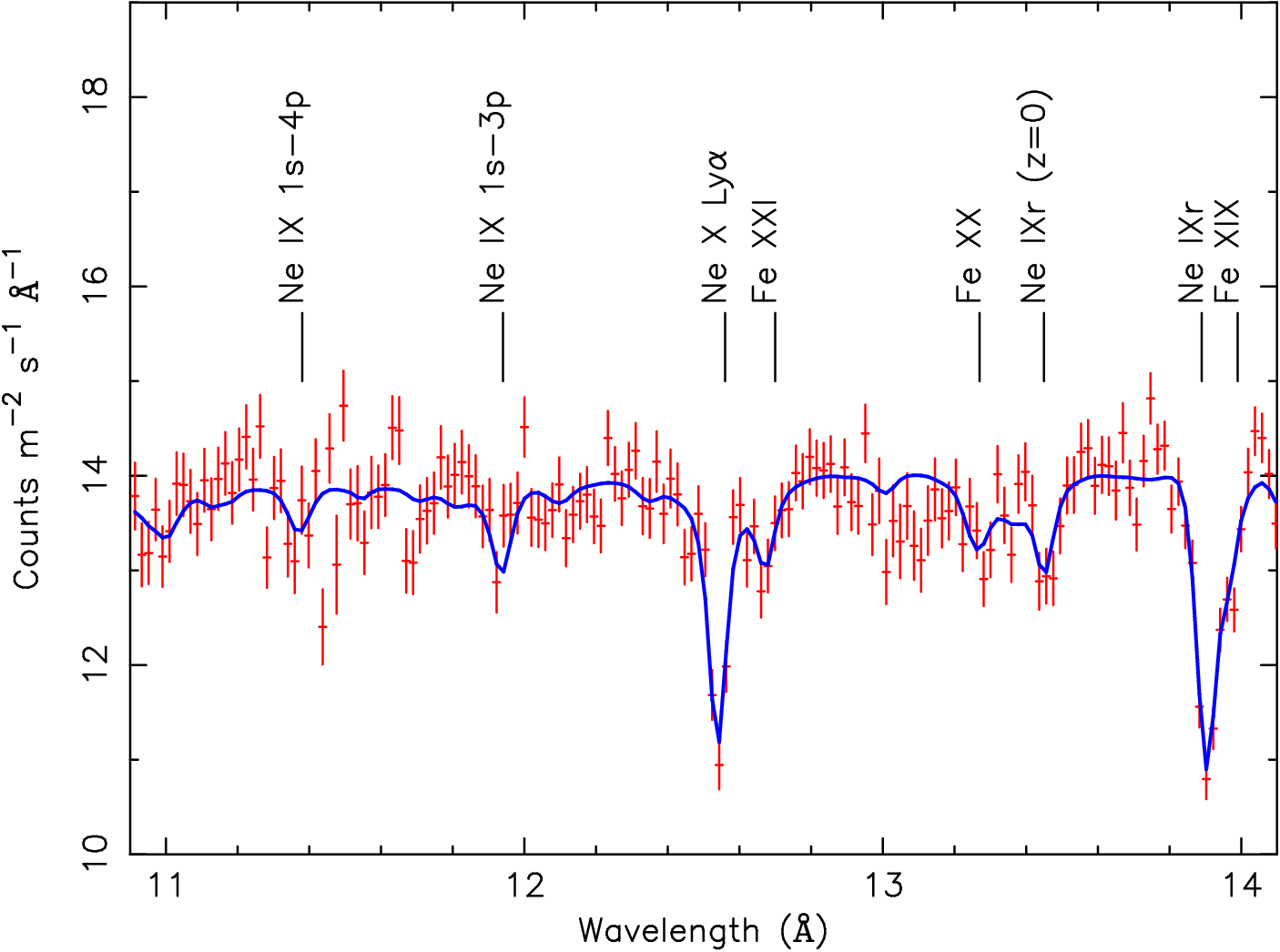}
\caption{\label{fig:spec1}
The data and the best fit to the RGS spectrum (model 2). The wavelengths shown are the observed wavelengths.}
\end{minipage}
\end{figure*}

\begin{figure*}[htbp]
\begin{minipage}[b]{1.0\linewidth}
\centering
\includegraphics[width=12cm, angle = -90]{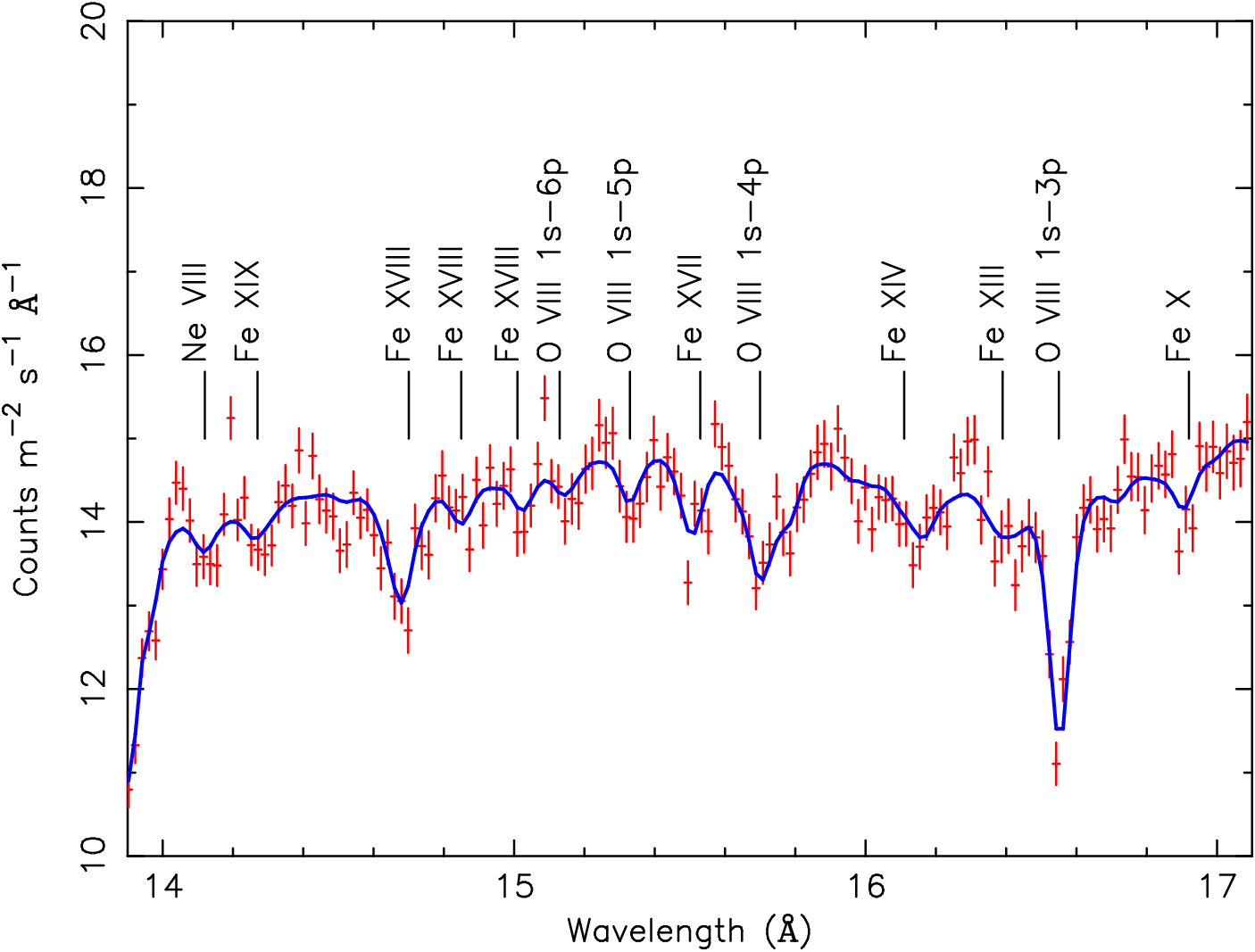}
\end{minipage}
\hspace{2cm}
\begin{minipage}[b]{1.0\linewidth}
\centering
\includegraphics[width=12cm, angle = -90]{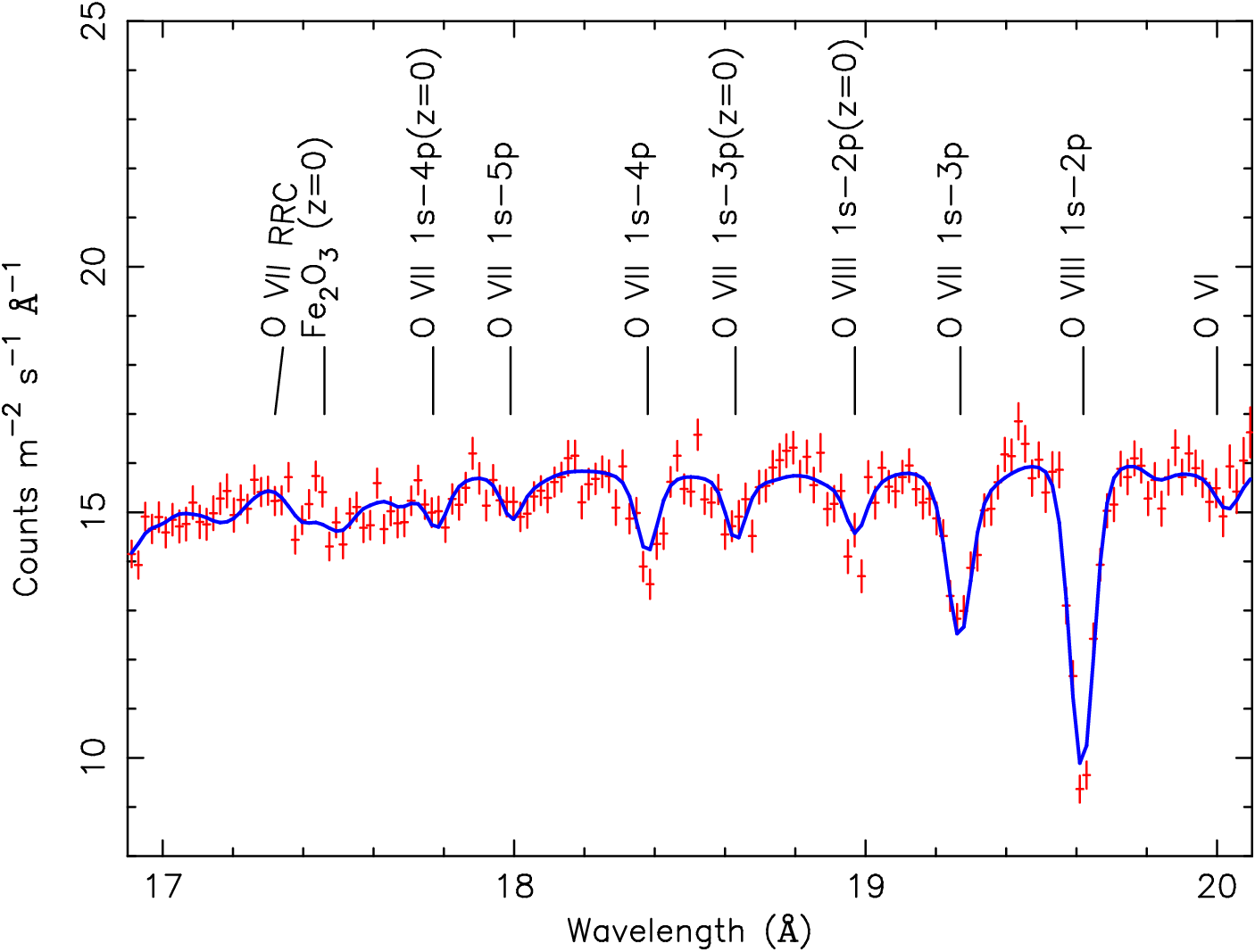}
\caption{\label{fig:spec2}
RGS spectrum continued.}
\end{minipage}
\end{figure*}

\begin{figure*}[htbp]
\begin{minipage}[b]{1.0\linewidth}
\centering
\includegraphics[width=12cm, angle = -90]{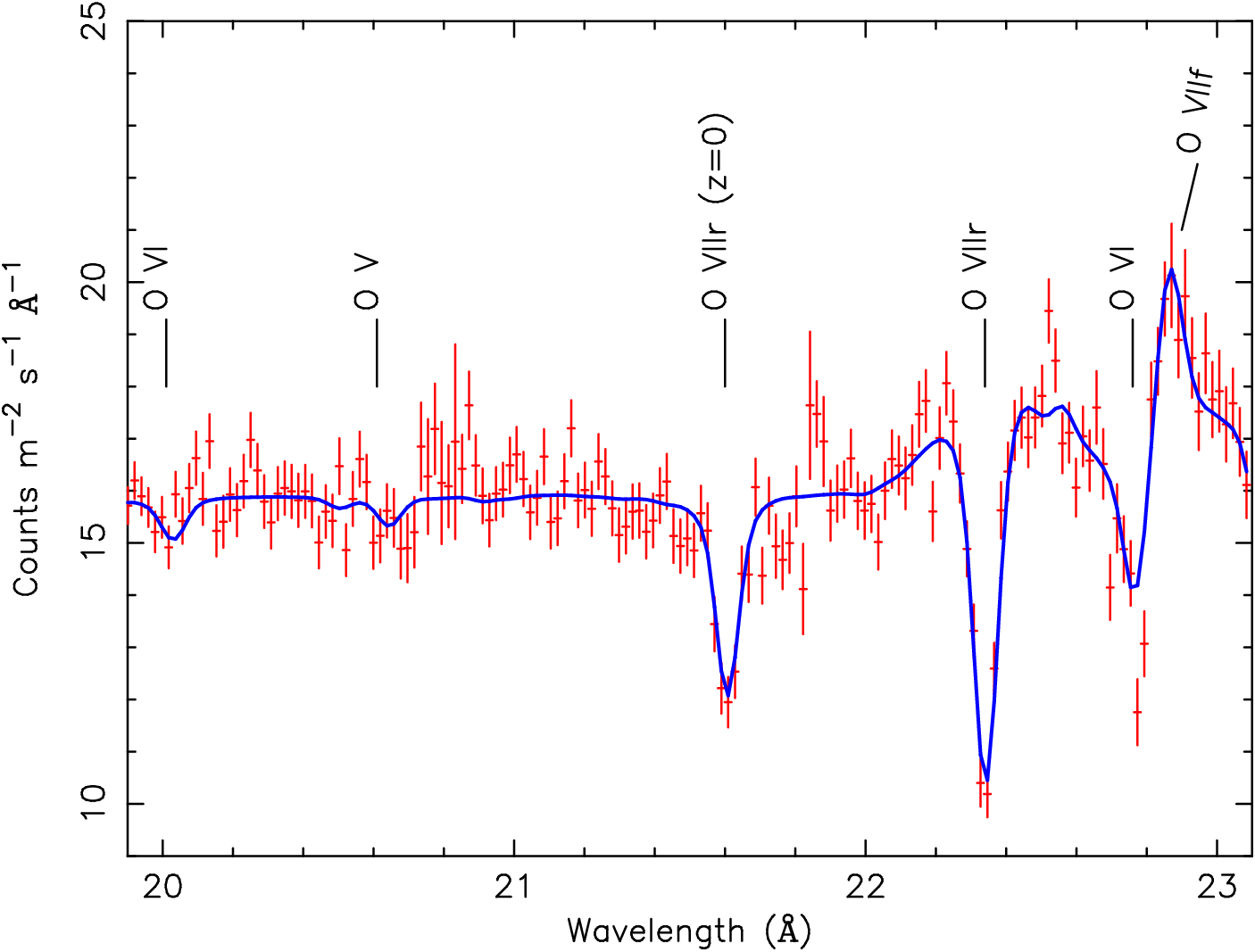}
\end{minipage}
\hspace{2cm}
\begin{minipage}[b]{1.0\linewidth}
\centering
\includegraphics[width=12cm, angle = -90]{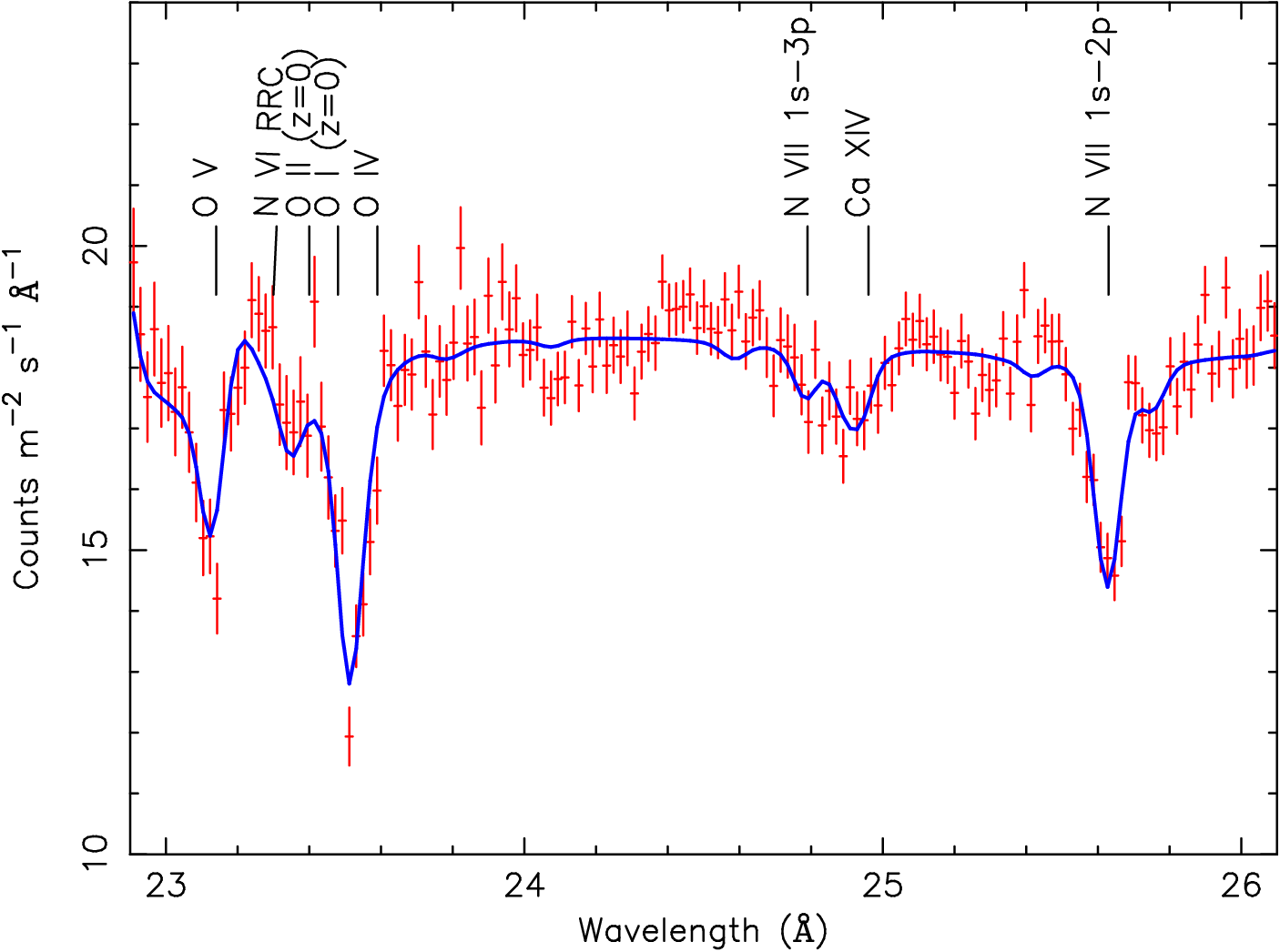}
\caption{\label{fig:spec3}
RGS spectrum continued.}
\end{minipage}
\end{figure*}

\begin{figure*}[htbp]
\begin{minipage}[b]{1.0\linewidth}
\centering
\includegraphics[width=12cm, angle = -90]{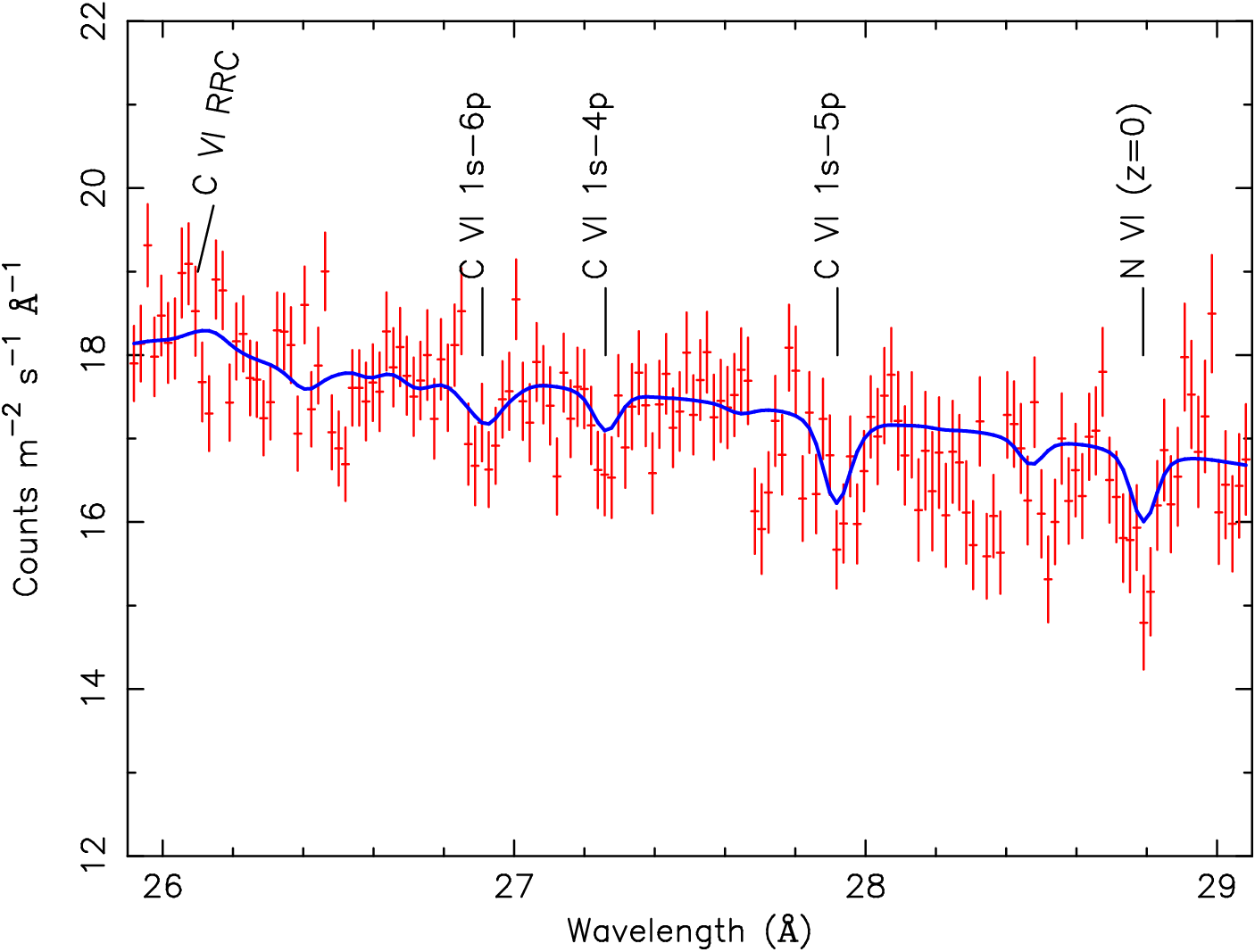}
\end{minipage}
\hspace{2cm}
\begin{minipage}[b]{1.0\linewidth}
\centering
\includegraphics[width=12cm, angle = -90]{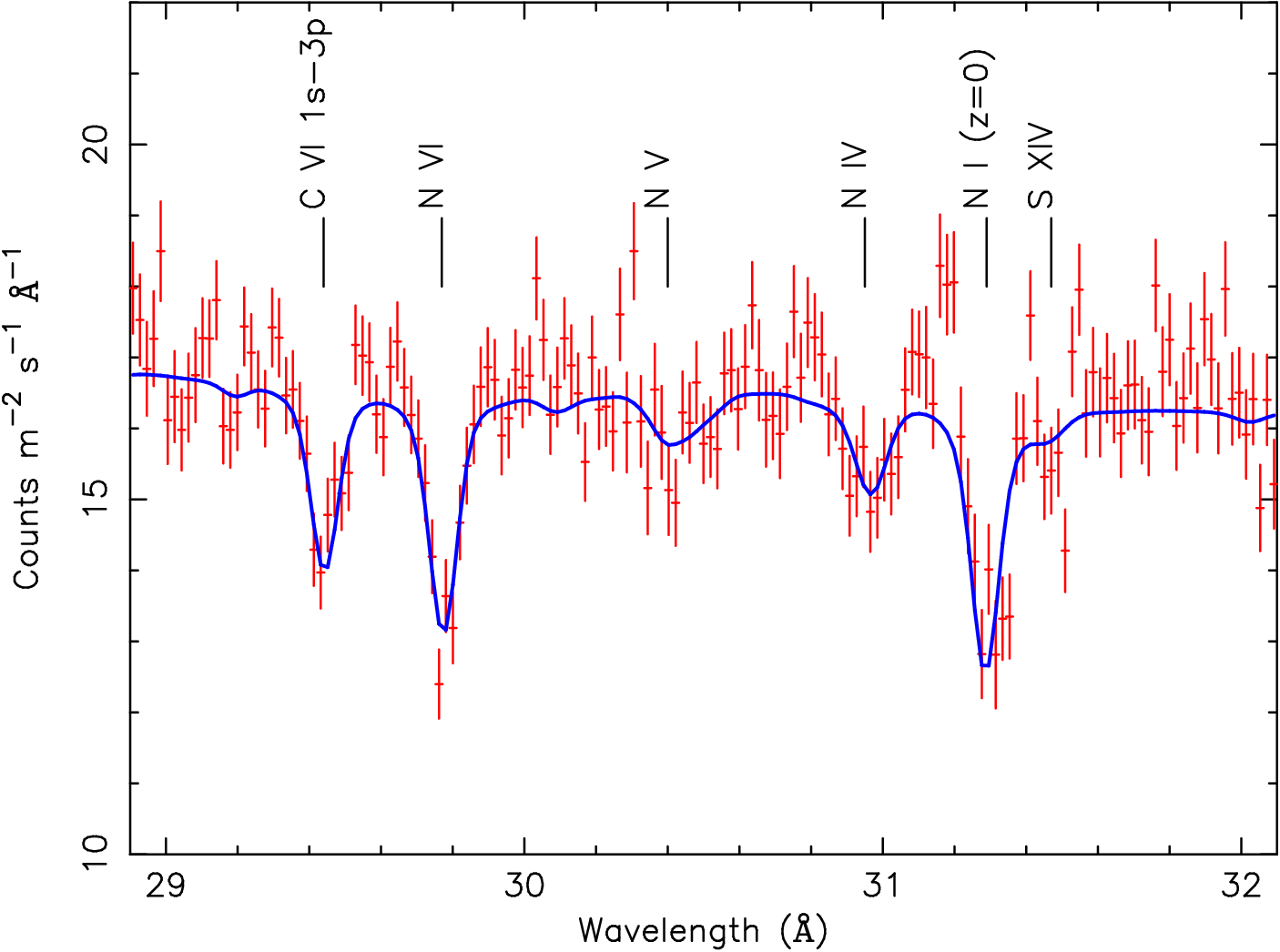}
\caption{\label{fig:spec4}
RGS spectrum continued.}
\end{minipage}
\end{figure*}

\begin{figure*}[htbp]
\begin{minipage}[b]{1.0\linewidth}
\centering
\includegraphics[width=12cm, angle = -90]{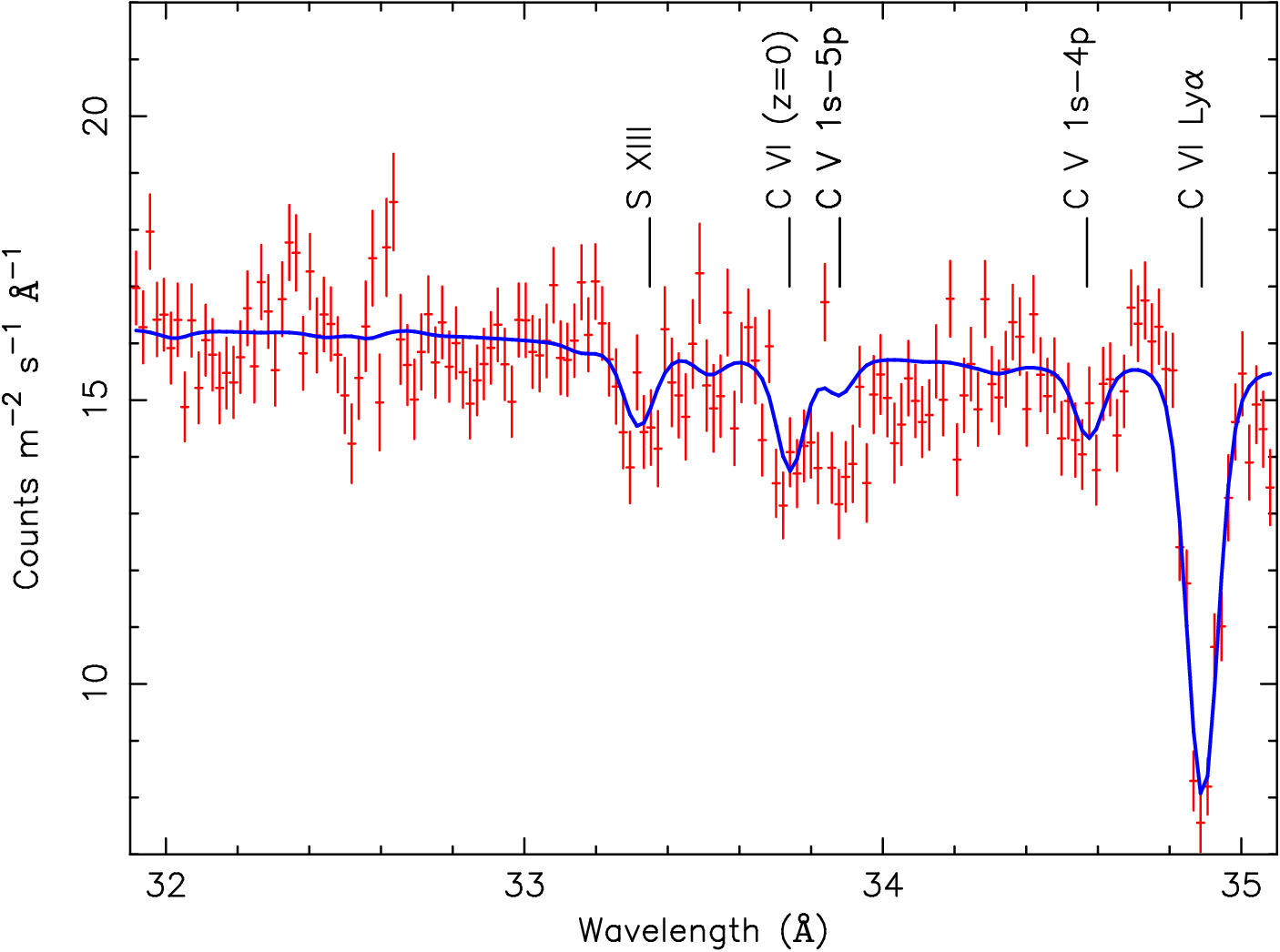}
\end{minipage}
\hspace{2cm}
\begin{minipage}[b]{1.0\linewidth}
\centering
\includegraphics[width=12cm, angle = -90]{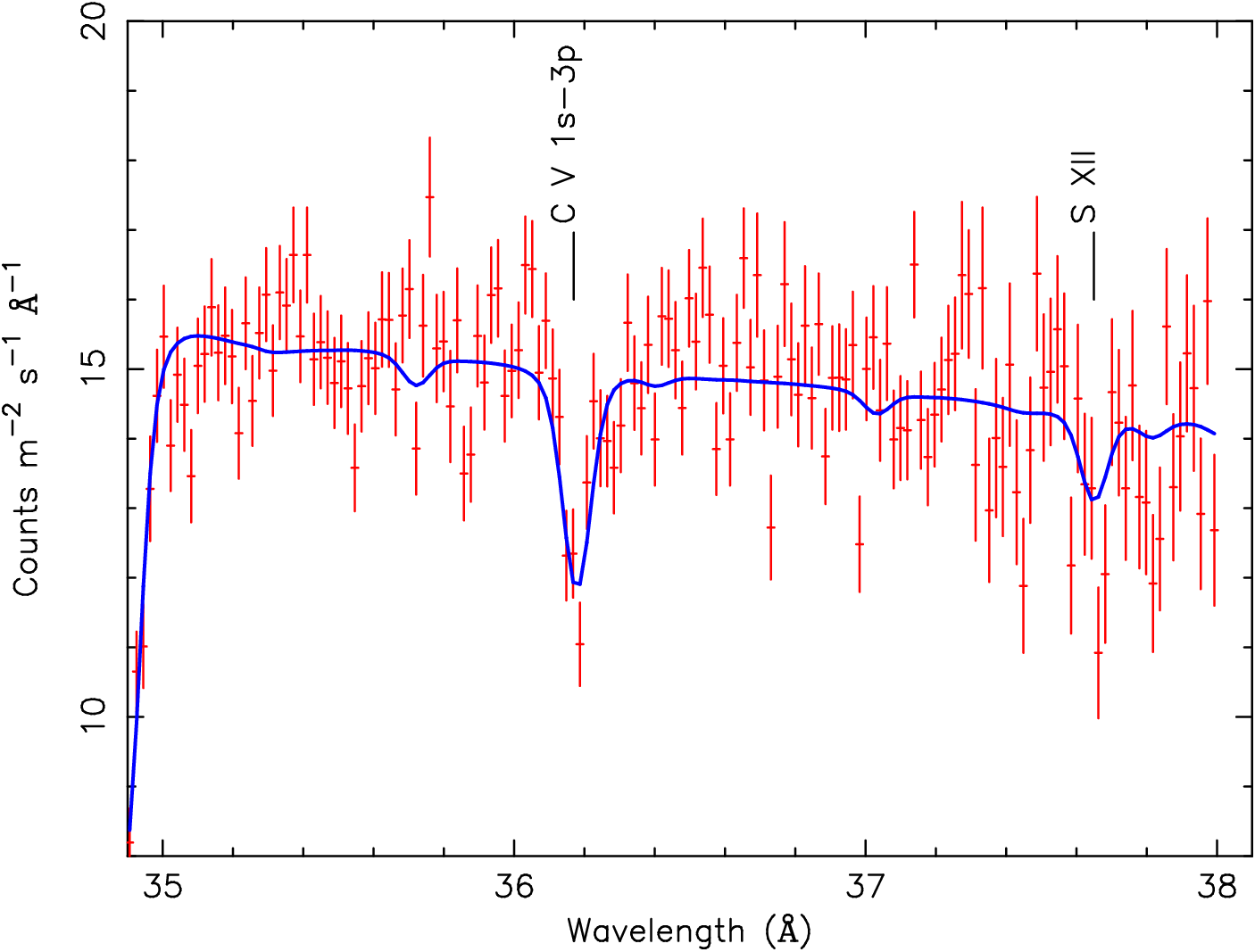}
\caption{\label{fig:spec5}
RGS spectrum continued.}
\end{minipage}
\end{figure*}

The RGS spectrum and the best$-$fit model (Model 2) are shown in Figs. \ref{fig:spec1} $-$ \ref{fig:spec5}. All the strongest absorption lines are labeled. Galactic lines are indicated with $z$ = 0. The spectrum has been re-binned to 0.02 $\AA$ bins for clarity. 
The model reproduces the data very well. We detect the \ion{O}{viii} Lyman series and the \ion{O}{vii} resonance transitions up to the 1s$-$5p transition, as well as the \ion{C}{vi} Lyman series up to the 1s$-$6p transition. We also detect the \ion{O}{vii} and \ion{O}{viii} series from the local $z$ = 0 component. The only features that are not reproduced well are the \ion{N}{vi} ISM absorption line at 28.78 $\AA$, a feature around 32.5 $\AA$, and another near the \ion{C}{v} absorption line at 33.9 $\AA$. The 32.5 $\AA$ feature is most likely due to small residuals in the RGS calibration, because it is much sharper and narrower than the other emission / absorption features. Its wavelength does not correspond to known major transitions. Also the \ion{O}{vi} absorption line at 22.8 $\AA$ is not well-fitted, possibly due to blending with the \ion{O}{vii} f narrow emission line. 

\subsection{Absorption measure distribution (AMD)}		\label{amd}

There has been a debate in the literature about whether the absorption measure
distribution defined here as $A(\xi) \equiv {\mathrm d}N_{\rm H}/{\mathrm
d}\xi$ is a smooth distribution spanning several decades in $\xi$ \citep[see e.g.][]{
Steenbrugge05} or consists of a limited number of discrete
components \citep[see e.g.][]{Costantini07a,Krongold03}.

We tested both alternatives as follows. We considered the total ionic
column densities derived from Model 1 (see Table \ref{tab:slab}), regardless of their
velocity, and fit them to a model using discrete components and to a model
with a continuous distribution. We simultaneously solved for the abundances of
the elements.

For Model 1, there is some arbitrariness in the assignment of ions to the two velocity components, in particular near the
ionization parameter $\log\xi = 2.1$ at the division, as for those ions both
velocity components will contribute. Thus the observed column densities near
that division are higher than the column density of a single velocity component.
This could result in the introduction of spurious ionization components or artificially
enhanced abundances. We tried to also do this analysis with the results for each velocity component
separately, but that is problematic. For Model 2, the error bars on the column densities for
individual velocity components are relatively high, because RGS
only partially resolves the lines of each component. This then gives too much
uncertainty to deduce conclusive results.

The first model we tested is a discrete model:
\begin{equation}
A(\xi)=\sum\limits_{i=1}^m H_i \delta(\xi - \xi_i),
\label{eqn:amd-discrete}
\end{equation}
Where $H_i$ are the total hydrogen column densities of the $m$ components with
ionization parameter $\xi_i$. 

From our runs with Cloudy we obtain for each ion $j$ curves for
the ion concentration $f_j(\xi)$ relative to hydrogen as a function of $\xi$,
assuming \citet{Lodders09} abundances. Given a set of values for $\xi_i$ and
$H_i$, it is then straightforward to predict the ionic column densities $N_j$:

\begin{equation}
N_j = \int\limits_0^{\infty} A(\xi) B_j f_j(\xi){\rm d}\xi,
\end{equation}
with $B_j$ the abundance in solar units of the parent element of ion $j$. We
solve this system by searching grids of models for different values of $\xi_i$,
and determine the corresponding best-fit column densities $H_i$ from a
least-squares fit to the data. The abundances are solved iteratively. We start
with solar abundances and solve for $H_i$. Then for each element we determine
its best-fit abundance from a least squares fit of its ionic column densities to
the predicted model of the last step. This procedure is repeated a few times
and converges rapidly. 

It should be noted that since we do not measure hydrogen lines, the hydrogen column
densities that we derive are nominal values based on the assumption of on
average solar metal$-$to$-$hydrogen abundance for the ions involved. In fact, we
derive only accurate \textit{relative} metal abundances. Truly absolute
abundances should be derived using UV data, but we defer the discussion on
abundances to later papers of this series \citep[][and Arav et al., in prep]{Steenbrugge11}.

It appears that we obtain the best solution if we take five ionization components
into account. Adding a sixth component does not improve the fit significantly, and
by deleting one, two or three components $\chi^2$ increases by 6, 16 and 200,
respectively (refitting in each case). Our best fit then has $\chi^2=42.5$ for 29 ions included in our
fit. The predicted model is shown in Table \ref{tab:slab}, together with the individual
contributions $\Delta\chi_j$ to $\chi^2$ for each ion (i.e.,
$\chi^2=\sum\Delta\chi_j^2$). Negative values for $\Delta\chi_j$ indicate lower observed ionic column densities than the model and positive values higher observed ionic columns. 
The best$-$fit parameters are shown in Table \ref{tab:robtab}.

\begin{table}
\caption{Parameters for the discrete distribution.}             % title of Table
\label{tab:robtab}      % is used to refer this table in the text
\centering                          % used for centering table
\begin{tabular}{l@{\,}c@{\,}c}        % centered columns (5 columns)
\hline\hline                 % inserts double horizontal lines
Component & $\log\xi^{a}$ & N$_{H}^{b}$ \\
\hline
A  & $-0.33\pm 0.49$  &  $0.23\pm 0.09$\\
B  & $ 0.71\pm 0.12$  &  $0.84\pm 0.10$\\
C  & $ 2.01\pm 0.02$  &  $4.8 \pm 0.4$\\
D  & $ 2.79\pm 0.06$  &  $5.7 \pm 0.9$\\
E  & $ 3.60\pm 0.27$  &  $54  \pm 73 $\\
\hline
\multicolumn{3}{l}{$^a$ Ionization parameter in 10$^{-9}$ W m.} \\
\multicolumn{3}{l}{$^b$ Column density in units of 10$^{24}$ m$^{-2}$.} \\
\end{tabular}
\end{table}

We did not include upper limits in our fit, and we also excluded the
argon lines because the predicted model is well below the marginal ``detections''
of \ion{Ar}{ix} and \ion{Ar}{x}. For further discussion, we also include
predicted column densities for hydrogen and \ion{C}{iii} and \ion{C}{iv},
although we cannot measure lines from these ions in the RGS band.

Next we consider a continuous AMD. It is impossible to make no a priori
assumptions for the shape of $A(\xi)$, but we minimize this as follows. We
assume that $\log A(\xi)$ is described by a cubic spline for $\log\xi$ between
$-3$ and $4$ with grid points separated by 0.2 in $\log\xi$. The use of
logarithms guarantees that $A(\xi)$ is non-negative; the spacing of $0.2$
corresponds to the typical scale on which ion concentrations change (making it
much smaller causes oversampling with unstable, oscillatory solutions), and the
range in $\xi$ covers the ions that are detected in the spectrum. Free
parameters of the model are the hydrogen column densities $H_i$ at the grid
points and the abundances. We solve for this system using a genetic algorithm
\citep{Charbonneau95}.

\begin{figure}[htbp]
   \includegraphics[angle= -90,width=9cm]{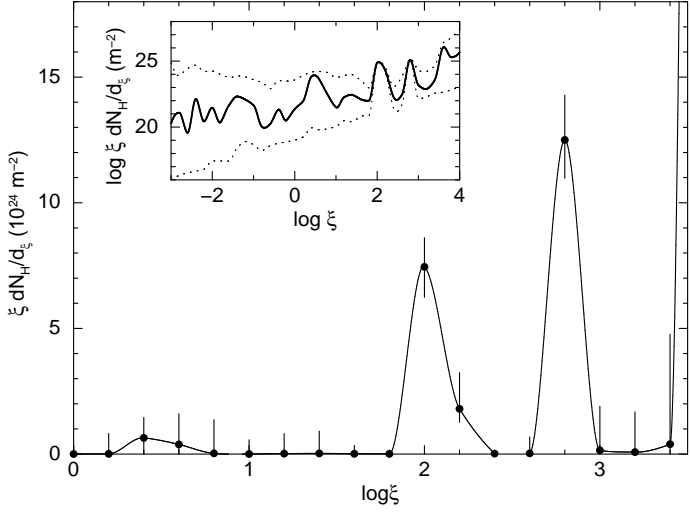}
   \caption{\label{fig:amd_model}
        Continuous absorption measure distribution plotted as $\xi A(\xi)$ for
Mrk~509. Shown is the median of the 117 runs with $\chi^2<\chi^2_{\min}+1$. Error
bars correspond to the minimum and maximum value of the AMD at each value of
$\xi$ for those runs with acceptable $\chi^2$. Note the two strong and isolated
peaks at $\log\xi = 2.0$ and $2.8$, respectively. The inset shows our results
on a logarithmic scale for a broader range of ionization parameter; dotted lines
connect the minimum and maximum values of all acceptable solutions.}
\end{figure}

We made 200 runs with the algorithm, and kept the 117 runs that resulted in
$\chi^2<\chi^2_{\min}+1$ with $\chi^2_{\min}=39.6$ the best solution. In
Fig.~\ref{fig:amd_model} we show the median of all these 117 good
solutions. The figure shows two strong, isolated peaks at $\log\xi = 2.0$ and $2.8$,
corresponding to components C and D of Table \ref{tab:robtab}. At a
higher ionization parameter ($\log\xi>3$), the range of component E of
Table \ref{tab:robtab}, there is also some AMD, but the detailed structure is essentially
unknown: there is a wide spread between the individual solutions that are
acceptable. At a lower ionization parameter ($\log\xi<1$), there is also some AMD
but again not a well-determined structure. A hint for the presence of component
B is that the median of the acceptable solutions is closer to the upper
limit in the range of $\xi$ between 0.4 $-$ 0.8.

As a final test, we extended the model with discrete components and
searched how broad the discrete components are. Replacing the $\delta$-function
by a Gaussian in (\ref{eqn:amd-discrete}), we get an upper limit to the $\sigma$
of the Gaussians of 0.06 and 0.13 in $\log\xi$ for the components C and D,
corresponding to a FWHM of 35 and 80\%. For the other components, there is no
useful constraint.

\section{Discussion}			\label{discussion}

\subsection{Foreground and emission features}

This paper focuses on the properties of the ionized outflow in Mrk 509, but given the quality of the data, a full description and discussion of all the features present in the spectrum is beyond the scope of this paper and will be given in a series of subsequent publications.
The only clear detections of narrow emission lines are the \ion{O}{vii} forbidden emission line at 22.101 $\AA$, the \ion{O}{vii} intercombination line at 21.802 $\AA$, and a narrow \ion{Ne}{ix} forbidden emission line at 13.70 $\AA$. The fluxes are consistent with earlier observations \citep{Detmers10}. We also searched for RRC features of the most prominent ions (C, N, and O) and found several possible weak RRCs. 
We do not detect any significant absorption due to neutral gas in the host galaxy. This means that we are observing the nucleus directly, which could indicate that we observe Mrk 509 almost face-on, as suggested in previous papers \citep{Phillips83,Kriss00,Kraemer03}.  

\subsection{Outflow models} \label{sect:outflow_model}

The warm absorber in Mrk 509 (\ion{O}{viii} column density of $\simeq$ 2 $\times$ 10$^{21}$ m$^{-2}$) is deeper than the one seen in Mrk 279 \citep[3 $\times$ 10$^{20}$ m$^{-2}$,][]{Costantini07a} but shallower with respect to those in NGC 5548 for instance \citep[3 $\times$10$^{22}$ m$^{-2}$, see e.g.][]{Steenbrugge05} or NGC 3783 \citep[4 $\times$ 10$^{22}$ m$^{-2}$, see e.g.][]{Behar03}. Nevertheless the high quality of this dataset allows for a thorough investigation of the outflow properties. 
While the main goal of the campaign is to localize the outflow, which requires investigating the ten individual observations, the integrated 600 ks spectrum is crucial for a full description of the properties (such as outflow velocity and ionization structure) of the outflow. To obtain the most accurate information about the true structure of the outflow we now compare the different models of the outflow. 

The first model (Model 1) is a very simple description, with only one velocity component for each ion. The velocity dispersion we obtain for the component that includes the \ion{O}{vii} and \ion{O}{viii} ions is 158 km s$^{-1}$. This is larger than what is obtained from the curve of growth analysis using a single velocity component for these ions \citep[96 and 113 km s$^{-1}$, respectively,][]{Kaastra2011b}. The reason for this difference is that having only one velocity component for these lines is an oversimplification. Adding a second velocity component for all ions (Model 2), improves the fit of the strong oxygen lines (\ion{O}{vii} and \ion{O}{viii}), with the sum of the velocity dispersions larger than for the single component case, but the total column density similar \citep{Kaastra2011b}. Thus the total ionic column densities for models 1 and 2 are consistent with each other. 

\begin{figure}[tbp]
\begin{minipage}[b]{1.0\linewidth}
\centering
\includegraphics[width=6cm, angle = -90]{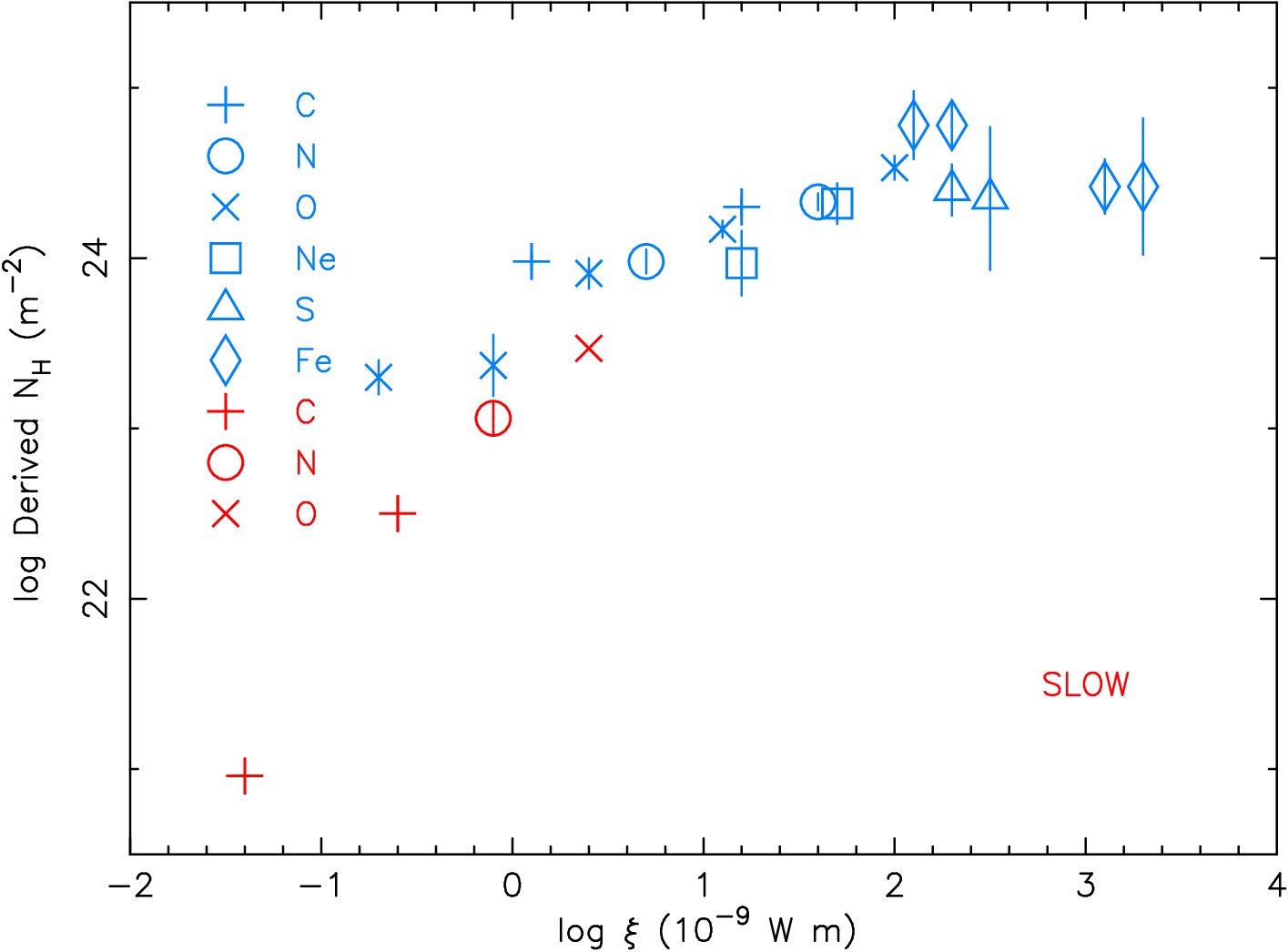}
\end{minipage}
\hspace{0.5cm}
\begin{minipage}[b]{1.0\linewidth}
\centering
\includegraphics[width=6cm, angle = -90]{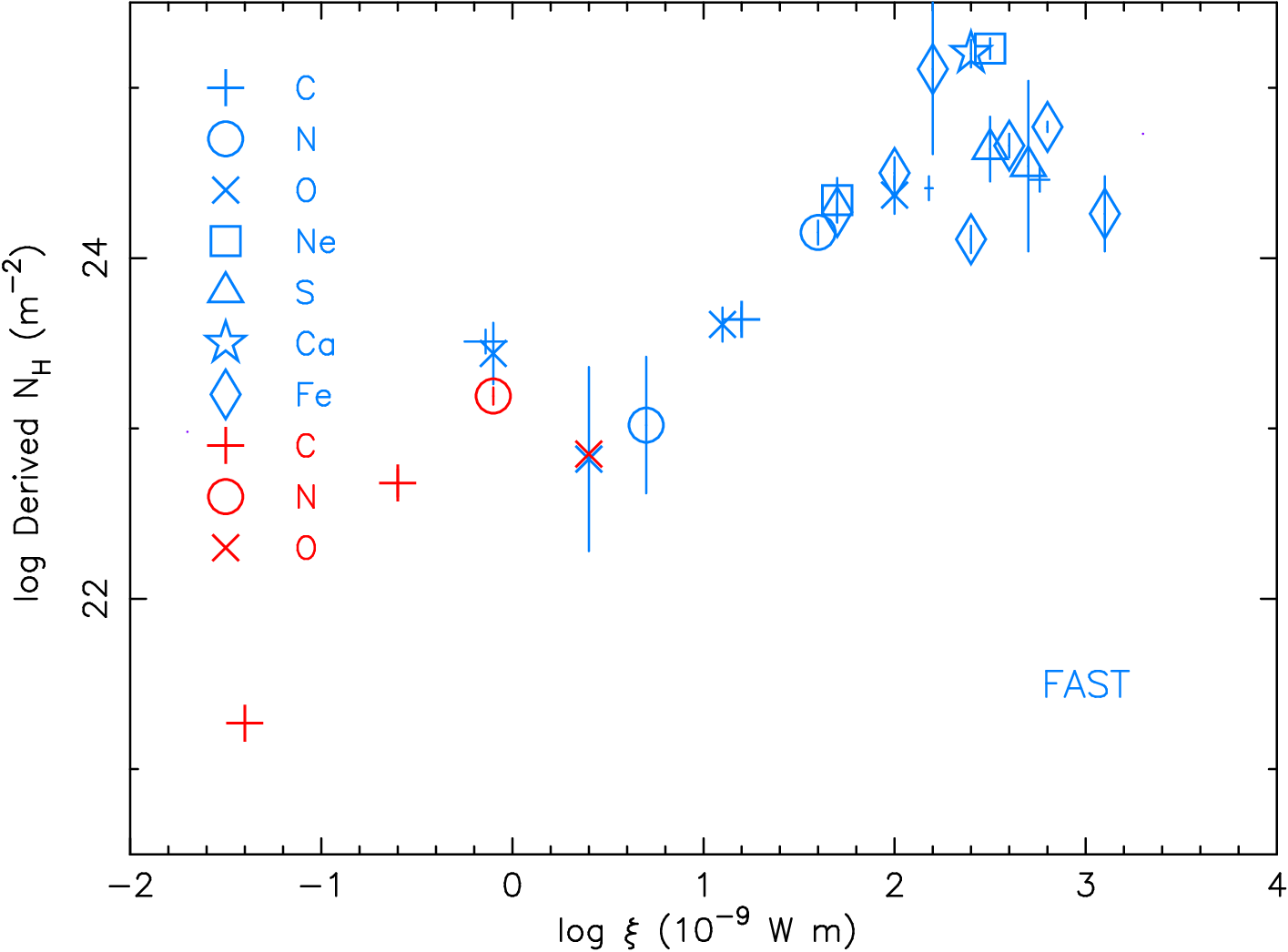}
\caption{\label{fig:amduv}
The derived hydrogen column density for every detected ion (see Table \ref{tab:slab}). We added archival UV data for \ion{C}{iii}, \ion{C}{iv}, \ion{N}{v}, and \ion{O}{vi} for comparison (shown in red). The top figure shows the distribution for the slow component, the bottom one shows the same for the fast component.}
\end{minipage}
\end{figure}

To compare the \textit{slab} models with model 3 we first need to convert the ionic column densities we measure into an equivalent hydrogen column density. There are two ways to do this. One is to assume that every ion occurs at the ionization parameter where its concentration peaks as a function of log $\xi$. This holds for some ions, but for others there is a wide range of ionization parameters where the ion makes a significant contribution. 
The alternative is that we take the full AMD method described in Sect. \ref{amd}. It is useful to compare these two methods so that we can see if there are major differences in the results and if these possible differences affect our conclusions. The results for the first method using Model 2 are shown in Fig. \ref{fig:amduv}, for both the slow and fast velocity components. Only ions for which we have a significant column density measurement are shown.
The results for the AMD method are shown in Fig. \ref{fig:amd_model}.   
What can be seen is that the AMD method clearly shows a discrete distribution of column density as a function of the ionization parameter. There is a clear minimum between the peaks at log $\xi$ = 2.0 and 2.8, where the column density is more than an order of magnitude less than at the two surrounding peaks. This indicates that there is almost no gas present at those intermediate ionization states. The simplified method (the one where we assume that every ion occurs at a single $\xi$ value) does show enhancements near the mean peaks of log $\xi$ = 2 and 2.8, but there are no clear minima in the distribution, although for the fast component there seems to be some hint for a minimum near log $\xi$ = 0.5. What is clear from this comparison is that the simplified method is unable to uncover essential details in the AMD. This is because not all ions are found at their peak ionization parameters. 
 
Another main difference between the models is that the \textit{slab} models (in contrast to Model 3) yield completely model-independent ionic column densities (i.e. no SED or ionization balance or abundances are assumed). This is an advantage if the atomic data for certain ions are uncertain, as the fit will not try to correct for this by changing the overall fit parameters or by poorly fitting this particular ion. From the measured ionic column densities we then can obtain the distribution of total hydrogen column density as a function of the ionization parameters. However at this step it requires the input of an ionization balance, hence an SED.

Model 3 is a direct fit using the ionization balance to predict the ionic column densities. The advantage of this model compared to Models 1 and 2 is that all ions, including those with a small column density are taken into account, and all the ionic column densities are connected through a physical model. There are fewer free parameters, so in principle a more accurate determination of the following parameters: $N_{H}$, $\xi$, $\sigma$ and $v$ can be obtained. 
The AMD method is a good combination of both models because the ionic column densities are determined model independently by the \textit{slab} models. Then the ionization balance is used to produce an AMD and obtain the number and parameters of \textit{xabs} components that are needed to describe the data properly. 

\begin{figure}[htbp]
   \includegraphics[angle= -90,width=9cm]{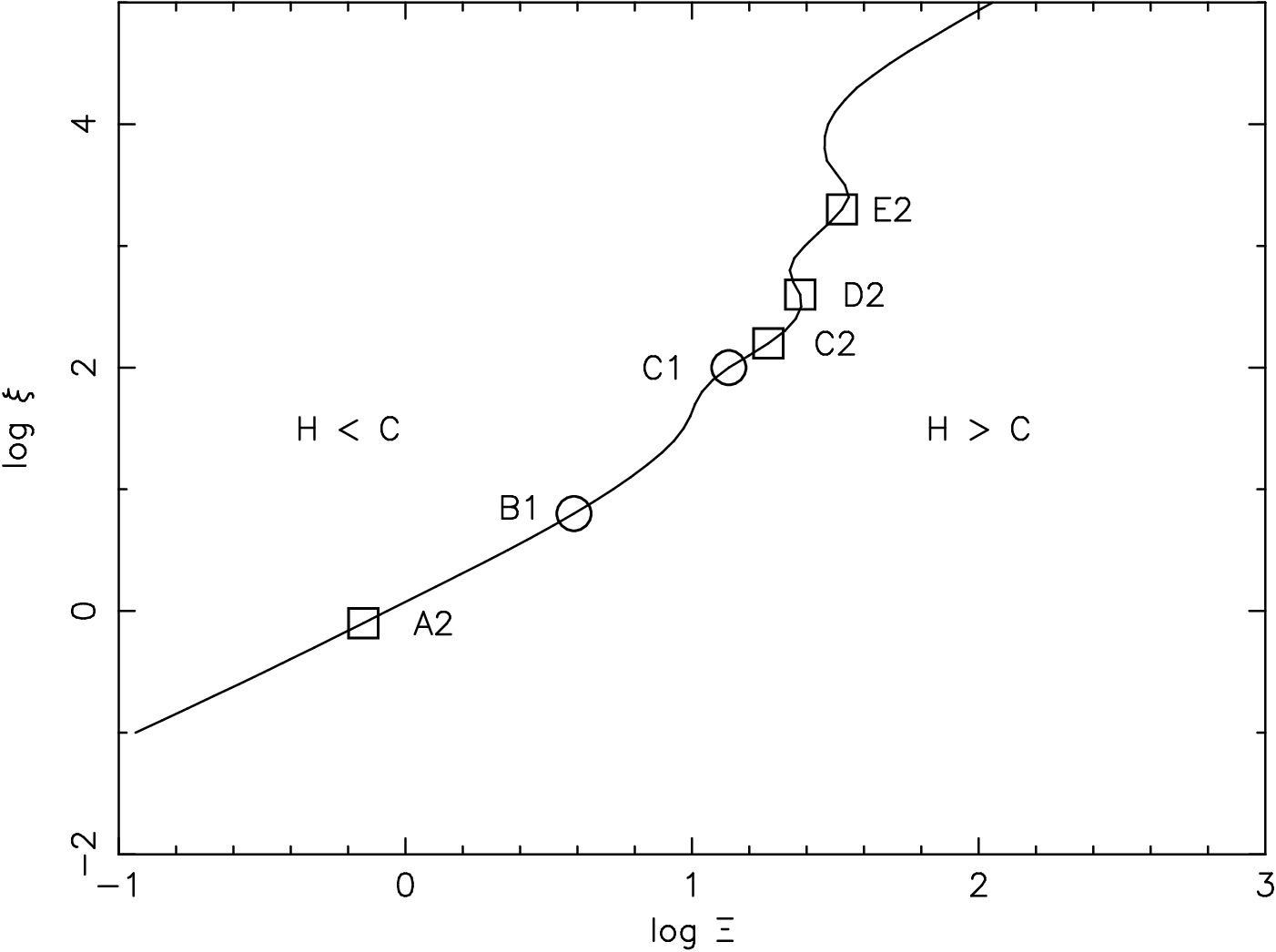}
   \caption{\label{fig:coolxabs}
        The cooling curve derived for the SED with the ionization parameters obtained from Model 3 over-plotted. The circles indicate the ionization parameters for the slow velocity component, while the squares shown are for the fast outflow. Regions where heating (H) or cooling (C) dominate are also indicated.}
\end{figure}

Model 3 and the AMD method make use of the ionization balance as determined from the SED. The resulting stability curve for the photoionized gas is shown in Fig. \ref{fig:coolxabs}. Components with the same $\Xi$ (in units of ) are in pressure equilibrium. 
Here $\Xi$ is defined in the following way:
\begin{equation}		\label{eq:xip}
\centering
      \Xi = 9610 \times \xi/T ,
\end{equation}
where $T$ is the temperature and $\xi$ is the ionization parameter in 10$^{-9}$ W m.  
The sections of the curve with a negative slope are unstable to perturbations. Not all components appear to be in pressure balance. For both velocity components the low-ionization gas (A2, B1) is not in pressure equilibrium with the higher ionization gas. This could indicate that the different gas phases are not colocated or that other forces (i.e. magnetic) are involved to maintain pressure equilibrium. 
It appears that most of the outflows in Seyfert 1 galaxies show gaps in the AMD \citep{Behar09}. This is probably due to (thermal) instabilities in the gas \citep{Holczer07}.
The exact nature of these apparent instabilities is still unclear (although a thermal scenario indeed seems plausible at the moment, based on the cooling curves, such as shown in Fig. \ref{fig:coolxabs}).

\subsection{Structure of the outflow}

Much work has already been done investigating the structure of the ionized outflows in other AGN \citep[see e.g.][for some examples]{Steenbrugge05,Holczer07,Costantini07a}. In most cases a wide range of ionization states has been detected, sketching the picture of a continuous distribution of the hydrogen column density as a function of $\xi$. However, there are also indications of a lack of ions in a certain temperature regime, where the ionized gas is in an unstable region of the cooling curve. In Mrk 509 such unstable regions occur for log $\xi$ between 2.4 and 2.8 and between 3.5 and 4. We have determined that the outflow in Mrk 509 is not continuous, but has discrete components, at least in the range of log $\xi$ = 2 $-$ 3. There are two main components, one at log $\xi$ = 2.0 and one at 2.8.
We also see a clear trend toward increasing column density for higher ionization states.    

We first discuss the ionization structure. The most pronounced component in our spectrum is component C (Table \ref{tab:robtab}).
The ionization parameter and total column density derived from our \textit{slab} fit (Table \ref{tab:robtab}), as well as the direct {\it xabs} fit (Table \ref{tab:xabs}), are fully consistent. From the analysis in Sect. \ref{amd} we find that this component is discrete and spans a very narrow range in ionization parameter: the FWHM is 35\%.  
Interestingly, according to our model (Table \ref{tab:slab}) component C contributes 50\% or more to the total ionic column density of 17 of the detected ions in our data set. Those ions span a range of $\log\xi = 1.15$ (\ion{O}{vii}) to 2.42 (\ion{S}{xiii}) in ionization parameter (see first column of Table \ref{tab:slab}).

The next most important component is component D at $\log\xi = 2.79$. It is responsible for the more highly ionized iron (up to \ion{Fe}{xx}) and the sulfur ions. It is mainly visible in the high$-$velocity component. Again the direct \textit{xabs} fit and the derived ionization parameter are fully consistent with each other, but the column density obtained from the \textit{xabs} fit is smaller by a factor of 3. This could be due to \textit{xabs} component C2 (log $\xi$ = 2.2), which also produces ions present in component D (log $\xi$ = 2.79). 
The other important component is B, which is responsible for most of the lower ionized carbon, nitrogen and oxygen ions. The ionization parameter and column density are fully consistent for both the derived model and the direct \textit{xabs} fit.  

Components A and E are also fully consistent with our Model 3, however they are only based on a few ions, so their exact column densities and ionization parameters are uncertain. Especially for the low-ionized gas, the lack of a strong UTA in Mrk 509 means that we only have upper limits on the column densities of the low ionized iron ions, up to \ion{Fe}{x}. This is also why the AMD distribution shown in Fig. \ref{fig:amd_model} has large uncertainties below log $\xi$ = 0.   

Earlier observations have detected only a few of the five components shown in Table \ref{tab:robtab}, owing to the poorer quality of the data. Using the Chandra HETGS, \citet{Yaqoob03} detected mainly component C, since the sensitivity of the HETGS is limited at longer wavelengths and component C is the strongest component.  
\citet{Smith07} detected mainly component B2, C (mixture C1 and C2), and D2 (possibly blended with E2). The outflow velocities in their analysis are different, as mentioned before, and they observe an inverse correlation between the outflow velocity and the ionization parameter. This analysis was based on the 2000 and 2001 archival data.  
\citet{Detmers10} analyzed earlier archival data (2005 and 2006) of Mrk 509 and found component B (possible mixture of the velocity components), C1, and D2 (possibly blended with E2). 
It is clear from these comparisons that, although all these earlier observations detected the main components of the outflow, in order to obtain a more complete picture of the outflow, we need a high$-$quality spectrum, like the one shown in this paper.  

Two of the three velocity components that we detected are consistent with earlier results, including the UV data \citep{Kriss00,Yaqoob03,Kraemer03}, and they correspond to the two main groups of UV velocity components, one at systemic velocity and the other at $-$ 370 km s$^{-1}$. Also in the X-ray regime, there is evidence of multiple ionization states for the same outflow velocity, such as components C2 and D2 in Table \ref{tab:xabs}. Also there are components that show a similar ionization state, but different outflow velocity, i.e. components C1 and C2 in Table \ref{tab:xabs}. Due to the almost zero outflow velocity of components B1 and C1, one could argue that these may be related to the ISM of the host galaxy. The UV spectra, with their much higher spectral resolution, can unravel the outflow, ISM, and redshifted high$-$velocity clouds \citep{Kriss11}.Generally speaking, the ionization parameter of the UV components is much lower than those of the X-ray components detected here. This could indicate that the UV and X-ray absorbers are cospatial, but have different densities. A full discussion of the connection between the UV and X-ray absorbers, as well as the geometry of the absorber, will be presented in \citet{Ebrero11}, where the simultaneous HST COS and Chandra LETGS data will be compared.  \\
We do not clearly detect the 200 km s$^{-1}$ redshifted component, which was found in the UV data \citep[velocity component 7 of ][]{Kriss00}. There is some indication that there could be an \ion{O}{vi} component at that velocity. There is some extra absorption at the red side of the line in Fig \ref{fig:spec3} at 22.78 $\AA$. We only obtain an upper limit for the \ion{O}{vi} column density in this velocity component of 10$^{20}$ m$^{-2}$. However, this is consistent with the lower limit from the UV data, which is 10$^{19}$ m$^{-2}$. 

\begin{figure}[tbp]
\begin{minipage}[b]{1.0\linewidth}
\centering
\includegraphics[width=7cm, angle = -90]{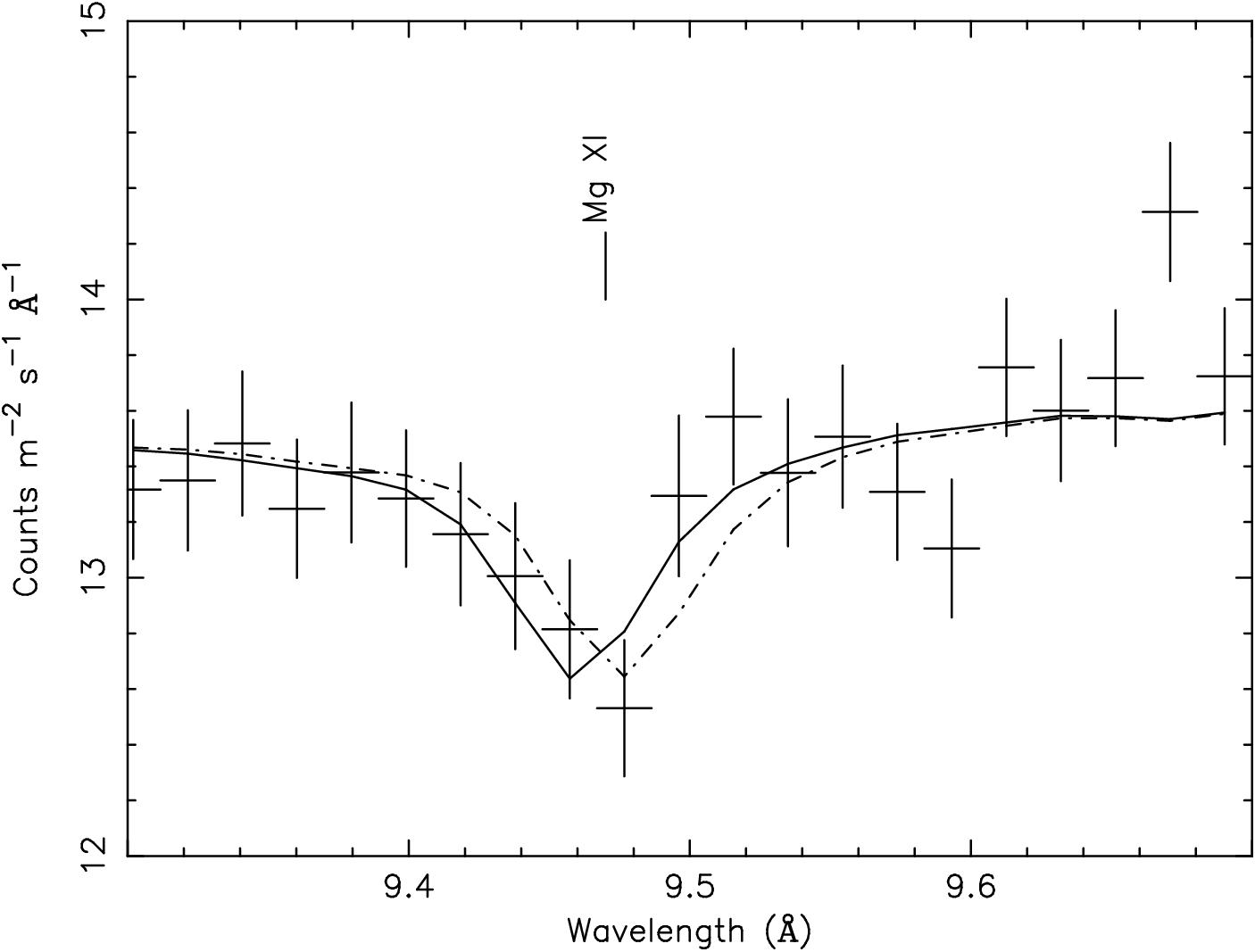}
\end{minipage}
\hspace{0.5cm}
\begin{minipage}[b]{1.0\linewidth}
\centering
\includegraphics[width=7cm, angle = -90]{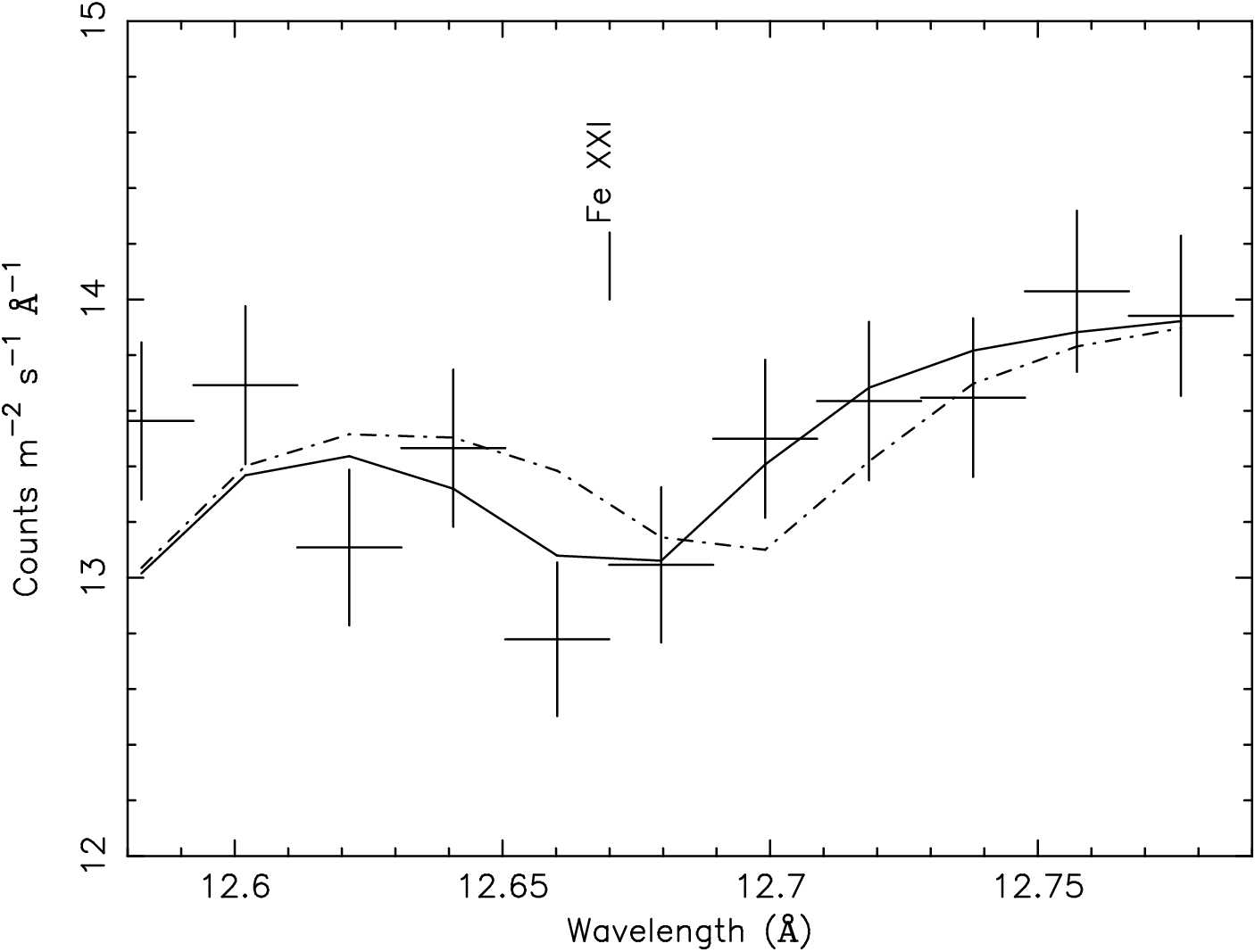}
\caption{\label{fig:highv}
The absorption lines of \ion{Mg}{xi} and \ion{Fe}{xxi}, which show a blueshift of 770 km s$^{-1}$. The dashed-dotted line shows a model with a blueshift of 300 km s$^{-1}$ and the solid line shows the model with a blueshift of 770 km s$^{-1}$.  }
\end{minipage}
\end{figure}

\begin{figure}[tbp]
   \includegraphics[angle= -90,width=9cm]{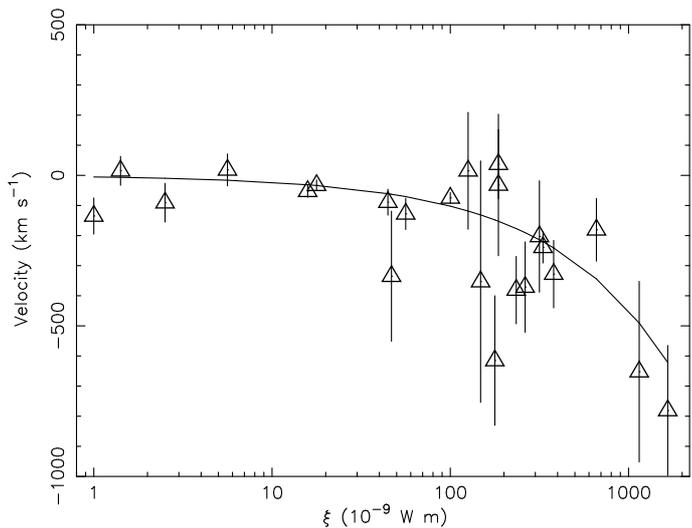}
   \caption{\label{fig:ionvelo}
        Outflow velocity vs. $\xi$. A general trend of increasing outflow velocity for ions with a higher ionization parameter can be seen. A negative velocity indicates outflow. }
\end{figure}

The highest velocity outflow component is only significantly detected in two ions (\ion{Mg}{xi} and \ion{Fe}{xxi}). Figure \ref{fig:highv} shows the two absorption lines fitted with a $-$ 770 km s$^{-1}$ velocity and a $-$ 300 km s$^{-1}$ velocity (just as component two in model 2). The improvement using the $-$ 770 km s$^{-1}$ component is $\Delta$$\chi^{2}$ = 16. We checked whether this component is also detected in other ions, but most of them (apart from \ion{Ne}{ix}) only yield upper limits to the ionic column density. We checked that the line profiles are the same for the separate RGS 1 and RGS 2 spectra and also for the first and second order spectra. In all cases the line profile is consistent with a 770 km s$^{-1}$ blueshift. Also a possible \ion{Mg}{xi} forbidden emission line cannot play a role here, because it is too far away (about 0.3 $\AA$) to influence the line profile in any way. This velocity component is consistent with an earlier Chandra HETGS observation, where there was an indication of this velocity component \citep{Yaqoob03}. A proper explanation of why this component is only clearly detected in these two ions and not in other ions with a similar ionization parameter is currently lacking. 
A trend visible in Table \ref{tab:slab} and Fig. \ref{fig:ionvelo} is that the higher ionized ions have a higher outflow velocity. Fitting a constant outflow velocity to the data yields a value of 70 $\pm$ 9 km s$^{-1}$ with $\chi^{2}$ = 72 for 24 d.o.f. A linear fit to the points gives a slope of 0.62 $\pm$ 0.07 and improves the $\chi^{2}$ to 46 for 24 d.o.f. If we instead fit a powerlaw, a relation of $v$ $\simeq$ $\xi^{0.64 \pm 0.10}$ is obtained, with a total $\chi^{2}$ of 34 for 24 d.o.f. The MHD models of \citet{Fukumura10} predict $v$ $\simeq$ $\xi^{0.5}$, which is consistent with the relation found here. However, it has to be noted that due to the blending of multiple velocity components (including gas that might not be outflowing at all), additional uncertainties are introduced that could affect the results. selecting only those ions, which clearly show blueshifts as well as more accurate outflow velocities, would be needed to investigate this trend further. 
The \ion{O}{iv} ion shows a large redshifted velocity, undetected in the other ions. Most likely this is due to the blending of the \ion{O}{iv} absorption line with the \ion{O}{i} line from the Galactic ISM at $z$ = 0. This blend makes it difficult to determine the centroid of the \ion{O}{iv} line exactly \citep{Kaastra2011b}. 

\subsection{Density profile}

A recent study has used the observed AMD to construct the radial density profile of the outflow in a number of sources \citep{Behar09}. Such an analysis is justified as long as the AMD is a smooth, continuous function of $\xi$, with the possible exception of unstable branches of the cooling curve, where gas may disappear to cooler or hotter phases. However in our case such an analysis is not justified, at least not for the range of log $\xi$ between $\sim$ 2 and 3 (components C and D). Here we clearly see narrow peaks in the AMD. At least for these components, this hints at rather localized regions with a limited density range, rather than to a large$-$scale outflow. 
At lower ionization parameters (components A and B), we cannot exclude a continuous distribution, owing to the limitations imposed by the line detection from the relevant ions. For these components, the situation is more complex due to the presence of both higher and lower velocity gas. Similarly, based on our analysis we cannot distinguish whether component E has a single component or a broader distribution on the second stable branch of the cooling curve (Fig. \ref{fig:coolxabs}). Gas on the third stable branch, at a very high ionization parameter, escapes our detection completely because of the lack of suitable lines in the RGS band. At best it could show strong lines from \ion{Fe}{xxv} or \ion{Fe}{xxvi} in the Fe$-$K band near 6.7 $-$ 7.0 keV, but the limited spectral resolution of EPIC combined with the likely moderate column densities prohibit us from detecting such a component in our data. 

\section{Conclusions} 		\label{conclusions}

We have presented one of the highest signal-to-noise RGS spectra of an AGN. With the almost unprecedented detail in this dataset, we could detect multiple absorption systems. The ionized absorber of Mrk 509 shows three velocity components, one at $-$13 $\pm$ 11 km s$^{-1}$, one at $-$319 km s$^{-1}$, and a tentative high$-$velocity component at $-$ 770 km s$^{-1}$. The first two components are consistent with the main absorption troughs in the UV. 
Thanks to the high$-$quality spectrum and the accurate column densities obtained for all ions, for the first time it has been shown clearly that the outflow in Mrk 509 in the important range of $\log\xi$ between 1$-$3 cannot be described by a smooth, continuous absorption measure distribution, but instead shows two strong, discrete peaks. At the highest and lowest ionization parameters, we cannot distinguish between smooth and discrete components. We also have found indications of an increasing outflow velocity versus ionization parameter.  
Large, dedicated multiwavelength campaigns such as this are the way forward, as this is currently the best method to investigate and characterize the outflows in the local Seyfert galaxies. 
 
\begin{acknowledgements}
This work is based on observations with \textit{XMM-Newton}, an ESA science mission with instruments and contributions directly funded by ESA Member States and the USA (NASA). SRON is supported financially by NWO, the Netherlands Organization for Scientific Research. KCS thanks ESO for its hospitality during part of this project and acknowledges the support of ComitŽ Mixto ESO - Gobierno de Chile.
Missagh Mehdipour acknowledges the support of a PhD studentship awarded by the UK Science \& Technology Facilities Council (STFC). P.-O.Petrucci acknowledges financial support from the CNES and the French GDR PCHE. M.Cappi, S. Bianchi, and G. Ponti acknowledge financial support from contract ASI-INAF n. I/088/06/0. N. Arav and G. Kriss gratefully acknowledge support from NASA/XMM-Newton Guest Investigator grant NNX09AR01G. Support for HST Program number 12022 was provided by NASA through grants from the Space Telescope Science Institute, which is operated by the Association of Universities for Research in Astronomy, Inc., under NASA contract NAS5-26555. E. Behar was supported by a grant from the ISF. G. Ponti acknowledges support via an EU Marie Curie Intra-European Fellowship under contract no. FP7-PEOPLE-2009-IEF-254279. 
\end{acknowledgements}

\newpage
\bibliographystyle{aa}
\bibliography{bibfiles}

\appendix

\section{Improving the atomic data}		\label{sect:app}

Due to the high quality of the Mrk 509 dataset, we have also updated some of the laboratory wavelengths of the important ions detected in the Mrk 509 spectrum. Table \ref{tab:atomic} shows all the updated lines. The \ion{O}{iii} lines are not resolved, so the strongest line at 23.071 was used \citep{Gu06} and the other two lines were shifted by the same amount. 
\begin{table}
\caption{Updated line list for important ions in the Mrk 509 spectrum.}             % title of Table
\label{tab:atomic}      % is used to refer this table in the text
\centering                          % used for centering table
\begin{tabular}{l@{\,}c@{\,}c@{\,}c}        % centered columns (4 columns)
\hline\hline                 % inserts double horizontal lines
Ion & Wavelength (new)  & Wavelength (old)  & Reference$^{1}$ \\ % table heading 
       & ($\AA$)                     &  ($\AA$)                  & \\
\hline
\ion{N}{vi}     & 28.7875 & 28.7870 & 1 \\
\ion{O}{iii}    & 22.9400 & 22.9784 & 2 \\
\ion{O}{iii}    & 23.0280 & 23.0489 & 2 \\
\ion{O}{iii}    & 23.0710 & 23.1092 & 2 \\
\ion{O}{v}      & 19.3570 & 19.3251 & 2 \\
\ion{O}{v}      & 19.9680 & 19.9242 & 2 \\
\ion{O}{vi}     & 18.2699 & 18.2896 & 2 \\
\ion{O}{vi}     & 18.2700 & 18.2897 & 2 \\
\ion{O}{vi}     & 18.5869 & 18.6059 & 2 \\
\ion{O}{vi}     & 18.5870 & 18.6060 & 2 \\
\ion{O}{vi}     & 19.1798 & 19.1355 & 2 \\
\ion{O}{vi}     & 19.1805 & 19.1362 & 2 \\
\ion{O}{vi}     & 19.3789 & 19.3412 & 2 \\
\ion{O}{vi}     & 19.3791 & 19.3414 & 2 \\
\ion{O}{vi}     & 22.0189 & 22.0063 & 3 \\
\ion{O}{vi}     & 22.0205 & 22.0079 & 3 \\
\ion{O}{vii}    & 17.7683 & 17.7680 & 1 \\
\ion{O}{vii}    & 18.6284 & 18.6288 & 1 \\
\ion{Ne}{viii}  & 13.6533 & 13.6460 & 4 \\
\ion{Ne}{viii}  & 13.6553 & 13.6480 & 4 \\
\ion{S}{xiii}   & 32.2380 & 32.2420 & 5 \\
\ion{S}{xiv}    & 30.4330 & 30.4270 & 5 \\
\ion{S}{xiv}    & 30.4750 & 30.4690 & 5 \\
\ion{Fe}{xvii}  & 15.2610 & 15.2650 & 6 \\ 
\ion{Fe}{xviii} & 14.3720 & 14.3780 & 7 \\
\ion{Fe}{xviii} & 14.5340 & 14.5400 & 7 \\
\ion{Fe}{xviii} & 14.5710 & 14.5550 & 7 \\
\ion{Fe}{xix}   & 13.4620 & 13.4650 & 7 \\
\ion{Fe}{xix}   & 13.5180 & 13.5210 & 7 \\
\ion{Fe}{xix}   & 13.7950 & 13.7980 & 7 \\
\ion{Fe}{xx}    & 12.8240 & 12.8130 & 7 \\ 
\ion{Fe}{xx}    & 12.8460 & 12.8270 & 7 \\ 
\ion{Fe}{xx}    & 12.8640 & 12.8470 & 7 \\ 
\ion{Fe}{xx}    & 12.9150 & 12.9040 & 7 \\ 
\ion{Fe}{xxi}   & 12.2840 & 12.2860 & 7 \\ 
\hline
\multicolumn{4}{l}{$^1$ References; (1) \citet{Engstrom95}; (2) \citet{Holczer10} ;}\\
\multicolumn{4}{l}{(3) \citet{Schmidt04}; (4) \citet{Peacock69};}\\
\multicolumn{4}{l}{(5) \citet{Lepson05}; (6) \citet{Brown98};}\\
\multicolumn{4}{l}{(7) \citet{Brown02}.}\\
\end{tabular}
\end{table}

\end{document}